\documentclass[%
 reprint,
 notitlepage,
superscriptaddress,
nofootinbib,
 amsmath,amssymb,
 onecolumn,
prd,
]{revtex4-1}

\usepackage{graphicx}
\usepackage{dcolumn}
\usepackage{bm}
\usepackage{hyperref}
\hypersetup{colorlinks}

\usepackage{color}
\usepackage[caption=false]{subfig}
\def\beq{\begin{eqnarray}}
\def\eeq{\end{eqnarray}}

\newcommand{\av}[1]{\langle{#1\rangle}} 
\let\vec\mathbf

\def\aap{\rm{A\&A}}

\def\apjl{\rm{ApJ}}                
\def\apjs{\rm{ApJS}}               

\newcommand{\hMpc}{h\,\mathrm{Mpc}^{-1}}
\newcommand{\inn}[2]{\left(\left.{#1}\right|{#2}\right)}

\definecolor{darkgreen}{RGB}{0,120,0}

\definecolor{brown}{RGB}{120,60,0}
\newcommand{\new}[1]{#1}
\newcommand{\resub}[1]{#1}

\begin{document}

\begin{flushright}
	CERN-TH-2020-146 \\
	INR-TH-2020-037
\end{flushright}


\title{Fewer Mocks and Less Noise:\\ \normalsize Reducing the Dimensionality of Cosmological Observables with Subspace Projections}

\author{Oliver H.\,E. Philcox}
\email{ohep2@cantab.ac.uk}
\affiliation{Department of Astrophysical Sciences, Princeton University,\\ Princeton, NJ 08540, USA}%
\affiliation{Department of Applied Mathematics and Theoretical Physics, University of Cambridge,\\ Cambridge CB3 0WA, UK}%
\affiliation{School of Natural Sciences, Institute for Advanced Study,\\ 1 Einstein Drive, Princeton, NJ 08540, USA}
\author{Mikhail M. Ivanov}
\affiliation{Center for Cosmology and Particle Physics, Department of Physics, New York University,\\ New York, NY 10003, USA}
\affiliation{Institute for Nuclear Research of the Russian Academy of Sciences,\\ 60th October Anniversary Prospect, 7a, 117312 Moscow, Russia}
\author{Matias Zaldarriaga}
\affiliation{School of Natural Sciences, Institute for Advanced Study,\\ 1 Einstein Drive, Princeton, NJ 08540, USA}
\author{Marko Simonovi\'c}
\affiliation{Theoretical Physics Department, CERN,\\ 1 Esplanade des Particules, Geneva 23, CH-1211, Switzerland}
\author{Marcel Schmittfull}
\affiliation{School of Natural Sciences, Institute for Advanced Study,\\ 1 Einstein Drive, Princeton, NJ 08540, USA}

\date{\today}

\begin{abstract}
Creating accurate and low-noise covariance matrices represents a formidable challenge in modern-day cosmology. We present a formalism to compress arbitrary observables into a small number of bins by \resub{projection} into a \resub{model-specific} subspace that minimizes the \resub{prior-averaged} log-likelihood error. The lower dimensionality leads to a dramatic reduction in covariance matrix noise, significantly reducing the number of mocks that need to be computed. Given a theory model, a set of priors, and a simple model of the covariance, our method works by using singular value decompositions to construct a basis for the observable that is close to Euclidean; by restricting to the first few basis vectors, we can capture almost all the \resub{constraining power} 
in a lower-dimensional subspace. Unlike conventional approaches, the method can be tailored for specific analyses and captures non-linearities that are not present in the Fisher matrix, ensuring that the full likelihood can be reproduced. The procedure is validated with full-shape analyses of power spectra from BOSS DR12 mock catalogs, showing that the 96-bin power spectra can be replaced by 12 subspace coefficients without biasing the output cosmology; this allows for accurate parameter inference using only $\sim 100$ mocks. Such decompositions facilitate accurate testing of power spectrum covariances; for the largest BOSS data chunk, we find that: (a) analytic covariances provide accurate models (with or without trispectrum terms); and (b) using the sample covariance from the MultiDark-Patchy mocks incurs a $\sim 0.5\sigma$ \resub{shift} in $\Omega_m$, unless the subspace projection is applied. The method is easily extended to higher order statistics; the $\sim 2000$-bin bispectrum can be compressed into only $\sim 10$ coefficients, allowing for accurate analyses using few mocks and without having to increase the bin sizes.
\end{abstract}

\maketitle


\section{Introduction}\label{sec: intro}


Most conventional analyses of cosmological data proceed by measuring a summary statistic, computing a theory model, and comparing the two in a Gaussian likelihood. A key ingredient in this is the inverted covariance matrix (also known as the precision matrix). In the simplest case, this is measured by creating a number of mock datasets, computing the desired statistic in each, then estimating the sample covariance directly. In order to obtain an unbiased precision matrix estimate, a large number of mocks is required \citep{2007A&A...464..399H}, and, it has been further shown that noise in the precision matrix leads to stochastic shifts in the best-fit parameters, \resub{which, is usually treated by inflating} the output parameter covariances \citep{2013PhRvD..88f3537D,2014MNRAS.442.2728T,2013MNRAS.432.1928T,2014MNRAS.439.2531P,2016MNRAS.456L.132S}. To reduce these \resub{shifts}, it is important to use a large number of mocks, though creating such a sample requires considerable computational effort, since the mocks are also required to be accurate. Furthermore, the magnitude of the effect increases with dimensionality; upcoming galaxy surveys such as those of DESI \citep{2013arXiv1308.0847L}, Euclid \citep{2011arXiv1110.3193L}, the Rubin Observatory  \citep{2019ApJ...873..111I}, and the Roman Telescope \citep{2015arXiv150303757S} will provide substantially higher resolution data, with an associated increase in number of bins.

There exists significant literature pertaining to alternative methods for galaxy survey covariance matrix generation. On one end of the scale sits purely theoretical models of the covariance, which, for the galaxy power spectrum and its multipoles, is possible via perturbation theory (PT) \citep{2016MNRAS.457.1577G,2019MNRAS.482.1786L,2020MNRAS.497.1684S,2019arXiv191002914W}. Whilst these are theoretically well motivated, we are fundamentally limited by the applicability of the perturbative model in the non-linear regime, and, without simulations, it is unclear how to set model hyperparameters such as galaxy bias and PT counterterms. A natural extension of this therefore is semi-analytic models, which combine well-understood theory and free parameters that can be calibrated from a (small) number of simulations. These exist for a range of statistics, including two-point correlation functions \citep{1994ApJ...424..569B,2016MNRAS.462.2681O,2019MNRAS.487.2701O,2020MNRAS.491.3290P,2019MNRAS.482.1786L}, three-point correlation functions \citep{2019MNRAS.490.5931P}, power spectra \citep{2016MNRAS.457..993P,2019MNRAS.482.4883C} and bispectra \citep{2019MNRAS.485.2806B}, though the applicability of the model assumptions must be rigorously tested in all circumstances.

An alternative approach is to find different ways of estimating the covariance matrix from simulations, for example by using noise reduction techniques (such as tapering \citep{2015MNRAS.454.4326P}, shrinkage \citep{2017MNRAS.466L..83J}, invoking sparsity \citep{2016MNRAS.460.1567P}, and \resub{combining with theoretical covariances \citep{2019MNRAS.483..189H}}), calibrating the covariance matrix with inexpensive small-volume simulations \citep{2017MNRAS.472.4935H,2018MNRAS.478.4602K}, or expanding the precision matrix around some smooth fiducial model \citep{2018MNRAS.473.4150F}. 
Of particular interest are schemes that reduce the dimensionality of the data-vector via some form of compression. As the number of bins falls, precision matrix noise becomes less important, allowing one to use less mocks to generate the covariance. The question, however, is how to perform this data compression.

In this work, we develop a method to compress a given data-vector 
by computing an information maximizing subspace from the theoretical model for our data, using techniques developed for gravitational wave analyses \citep{2019PhRvD..99l3022R}. In practice, we take a large sample of $N_\mathrm{bank}$ points in the underlying \resub{(prior-bounded)} space of $N_\mathrm{param}$ cosmological and nuisance parameters, and compute model data vectors (which we denote the `template bank') at each point. \resub{This requires no knowledge of the observational data-vector; just the theory model and parameter priors.} Using a singular value decomposition of the resulting $N_\text{bank}\times N_\text{bin}$ noise-weighted matrix, we generate a set of basis vectors which are ordered in signal-to-noise and, together, contain all cosmological information. These do not require prior knowledge of the observed data-vector. By restricting to the first $N_\mathrm{SV}$ of these modes, we capture the dominant contributors to the specific experimental likelihood, in a much lower dimensional space. The analysis then proceeds using the projection of the full data-vector into this subspace; \resub{we note that it is fully applicable to non-Gaussian posteriors (including multi-modality) and blind to the choice of input parameters.} Assuming one has a somewhat accurate initial ansatz for the covariance, this procedure \resub{captures the maximum log-likelihood information possible} 
for a fixed dimensional data vector using linear transformations. \resub{Furthermore, it is of low-cost, requiring only a set of theory model evaluates that could easily be evaluated at the start of an MCMC routine.}

Ours is certainly not the first work performing analysis along these lines, with other examples including bispectrum compression via approximate eigenvalue decompositions \citep{2000ApJ...544..597S} and expansions in polynomial basis functions \citep{2009PhRvD..80d3510F,2010PhRvD..82b3502F,2012PhRvD..86f3511F,2013PhRvD..88f3512S}, as well as the \texttt{COSEBI} method for cosmic shear observables \citep{2010A&A...520A.116S,2014MNRAS.440.1379E}. Perhaps most notable are the \texttt{MOPED} algorithm \citep{2000MNRAS.317..965H,2001MNRAS.327..849R,2017MNRAS.472.4244H,2020MNRAS.tmp.2512H} and Karhunen-Lo\`eve methods \citep{1997ApJ...480...22T}, both of which perform linear transformations to compress the data, \resub{with the former outputting} a set of $N_\mathrm{param}$ values which preserve the Fisher matrix. Whilst these have been shown to have utility in cosmological analyses \citep{2019PhRvD.100h3502P,2018MNRAS.476.4045G} most \resub{define the basis vectors} by assuming the posterior surface to be \resub{locally} Gaussian, \textit{i.e.}~that all information is contained within the second-derivatives of the log-likelihood, and hence the Fisher matrix. Further, the compression \resub{can be computationally expensive for high-dimensional data-sets and is optimal} only for the linear response of the model data vector with respect to parameter changes around a fiducial point in model space, without consideration of whether this is valid across the parameter space spanned by the priors. \resub{For most cosmological analyses, these issues do not pose a significant problem, especially if the procedure is iterated, but caution is needed in the case of highly non-linear posteriors.} \resub{Refs.\,\citep{2018MNRAS.476L..60A,2018PhRvD..97h3004C} provide an interesting generalization of these approaches to non-Gaussian likelihoods, via compression to a \textit{score function}.} The subspace method developed herein instead optimizes the log-likelihood itself, ensuring that the $\chi^2$ factors of the original and compressed likelihoods agree up to some fixed limit across the whole prior domain. \resub{Unlike approaches simply decomposing the covariance matrix of the statistic, our approach is tailored to each specific analysis, effectively ensuring that we do not encode information that is well known \textit{a priori}.}  Notably, this can require using a few \textit{more} modes than parameters to encapsulate non-linear parameter responses, though the precise number can be robustly set from accuracy considerations. Furthermore, whilst many methods use an approximate covariance of the compressed data-points (e.g., by assuming them to be independent) allowing for analytic posterior sampling, we instead opt to use the full covariance for accuracy, as in Refs.\,\citep{2017MNRAS.472.4244H,2018MNRAS.476.4045G}.

Whilst our approach is fully general and can be applied to any observable, given a theory model, parameter priors and a fiducial estimate of the covariance, we here consider the analysis of galaxy survey data in redshift-space, using the approach of Refs.\,\citep{2020JCAP...05..042I,2020PhRvD.101h3504I,2020JCAP...05..032P,2020arXiv200410607C,2020arXiv200308277N,2020arXiv200808084P}. By creating a template bank from the theory model, we can generate a subspace of much reduced dimension, which, for a large number of mocks, gives similarly accurate parameter inference to the full likelihood, and, for few mocks, substantially reduces the noise-induced \resub{shifts}, \resub{avoiding the need for significant posterior inflation}. 

The paper has the following structure. In Sec.\,\ref{sec: methodology}, we provide a mathematical derivation of the template bank and subspace decomposition, as well as discussing the corresponding subspace likelihoods. Sec.\,\ref{sec: shifts} discusses the expected constraints obtained from the subspace analysis and the effects of noise in the data and covariance. We apply the formalism to the galaxy power spectrum in Sec.\,\ref{sec: applications}, before discussing extensions to other analyses in Sec.\,\ref{sec: discussion} and concluding with a summary in Sec.\,\ref{sec: summary}. Appendices \ref{appen: fisher-matrices}\,\&\,\ref{appen: param-shifts} contain derivations of key results used in Sec.\,\ref{sec: shifts}, with Appendix \ref{appen: an-marg} presenting additional material related to Sec.\,\ref{sec: applications}.


\section{Methodology}\label{sec: methodology}

\subsection{The Template Bank and Euclidean Subspace}\label{subsec: template-bank}
A general cosmological model is specified by a set of parameters $\vec\theta$, which can be fundamental, nuisance or systematic. Whilst the arguments below strictly apply to any physical model, this work will principally be concerned with the analysis of galaxy power spectra, via the one-loop Effective Field Theory (EFT) model \citep{2020JCAP...05..042I,2020PhRvD.101h3504I,2020JCAP...05..032P,2020arXiv200410607C,2020arXiv200308277N,2020arXiv200808084P}. This carries the parameter vector
\beq\label{eq: parameter-space}
    \vec\theta = \{\omega_\mathrm{cdm}, A_s/A_{s,\mathrm{fid}}, h,...\}\times\{b_1,b_2,b_{G_2},b_4,c_{s,0},c_{s,2},P_\mathrm{shot}\},
\eeq
where additional parameters can be added to the first set to more finely probe cosmology. Here, we will use only these three for simplicity, but we note that the method can be arbitrarily extende. 

The model parameters generate a manifold to which we can associate a vector field of $N_\mathrm{bin}$-dimensional model spectra, $P_a(\vec\theta)$, where $a$ specifies the component ($k$-bin and multipole) \new{of the power spectrum evaluated at $\vec\theta$}.
We may define the inner product between two points \resub{$\vec\theta^{(i)}$ and $\vec\theta^{(j)}$} as the noise weighted distance from a suitably defined mean spectrum, $\overline{P}$, \new{(later be set to the mean power spectrum from the template bank)};
\beq
    \av{\delta P^{(i)},\delta P^{(j)}}_\mathsf{C} = \sum_{ab}\delta P_a^{(i)}\mathsf{C}^{-1}_{ab}\delta P_b^{(j)}
\eeq
where $\mathsf{C}$ is some fiducial covariance, encoding the metric, and $\delta P^{(i)} = P(\vec\theta^{(i)}) - \overline{P}$.\footnote{\resub{Note that this assumes the manifold to be Riemannian, such that the `distance' between points is specified by a quadratic form. Below, this will correspond to the assumption of a Gaussian noise-model.}} This is motivated by considerations of the $\chi^2$ of a sample power spectrum $\hat{P}$ with model $P(\vec\theta)$ and covariance $\mathsf{C}_D$, which is simply a squared norm;
\beq
    \hat{\chi}^2(\vec\theta) = \av{\hat{P}-P(\vec\theta), \hat{P}-P(\vec\theta)}_{\mathsf{C}_D}.
\eeq

It is convenient to perform a global linear transform that centers and whitens the power spectra,
\beq\label{eq: X-def}
    \delta P(\vec\theta)\rightarrow X(\vec\theta),\qquad X_a(\vec\theta) \equiv \sum_{ab}\mathsf{C}^{-1/2}_{b}\left[P_b(\vec\theta)-\overline{P}_b\right],
\eeq
\resub{(where $C^{1/2}$ is the Cholesky factorization of $C$)}, 
whilst preserving the inner product;
\beq\label{eq: inner-product-X}
    \av{\delta P^{(i)},\delta P^{(j)}}_\mathsf{C} \equiv \av{X^{(i)}|X^{(j)}} = \sum_{abcd}X_{a}^{(i)}\mathsf{C}^{1/2}_{ab}\mathsf{C}^{-1}_{bc}\mathsf{C}^{1/2}_{cd}X_d^{(j)} = \sum_{a}X^{(i)}_aX^{(j)}_a.
\eeq
Note that the metric is Euclidean in this set of co-ordinates.\footnote{\new{When computing the data $\chi^2$ we will later find that the metric becomes \textit{almost} Euclidean, providing the true and fiducial covariances are similar.}}
We can write an arbitrary rotated spectrum $X(\vec\theta)$ in terms of a set of $N_\mathrm{bin}$ orthonormal basis vectors $V_\alpha$, \textit{i.e.}
\beq\label{eq: basis-decomposition}
    X_a(\vec\theta) = \sum_{\alpha}c_\alpha(\vec\theta) V_{\alpha a}, \qquad c_\alpha(\vec\theta) = \sum_a X_a(\vec\theta)V_{\alpha a}
\eeq
with coefficients $c_\alpha$, where the basis vectors $V_\alpha$ satisfy $\av{V_\alpha|V_\beta} = \delta^K_{\alpha\beta}$. Hence the inner product of any two (rotated) vectors can be written as a product of basis coefficients;
\beq
    \av{X^{(i)}|X^{(j)}} = \sum_{\alpha}c^{(i)}_\alpha c^{(j)}_\alpha.
\eeq

To compute the basis vectors, $V_\alpha$, of Eq.\,\ref{eq: basis-decomposition}, we sample a set of $N_\mathrm{bank}$ spectra from the parameter manifold (known as the `template bank') and execute a singular value decomposition.\footnote{Note that we restrict the parameter ranges (\textit{i.e.} apply broad priors), such that the manifold is of finite extent.} Practically, this identifies the directions in the vector space which have the largest contributions to the log-likelihood, $\chi^2$. Given a set of points, $\{\vec\theta^{(i)}\}$, on the (bounded) manifold with corresponding spectra $\{P^{(i)}\}$ \new{(each of dimension $N_\mathrm{bin})$}, we first compute the rotated spectra $\{X^{(i)}\}$ at each point using Eq.\,\ref{eq: X-def}, setting $\overline{P}$ to the mean of all template bank spectra.\footnote{\new{Subtracting the mean is necessary to perform SVDs, but it does not change the analysis, since $\overline{P}$ cancels in the likelihood.}}
By stacking the rotated spectra, we can form an $N_\text{bank}\times N_\text{bin}$ matrix $X_{ia}$, which has the SVD,
\beq\label{eq: SVD}
    X_{ia} = \sum_\alpha U_{i\alpha} D_\alpha V_{\alpha a},
\eeq
where $D_\alpha$ is a \resub{rank-$N_\mathrm{bin}$} diagonal matrix of the singular values (SVs) and the matrices $U$ and $V$ \resub{(of dimension $N_\mathrm{bank}\times N_\mathrm{bin}$ and $N_\mathrm{bin}\times N_\mathrm{bin}$)} are unitary, projecting from observations to SVs, and SVs to spectra respectively.\footnote{Since the SVD algorithm is linear in the number of posterior samples, generating basis vectors is not a computationally intensive step.} By comparison with Eq.\,\ref{eq: basis-decomposition}, we see that the basis coefficients are given by
\beq
    c^{(i)}_\alpha = U^{(i)}_\alpha D_\alpha.
\eeq

The principal value of this decomposition is that any spectrum can be well approximated by a relatively small number of basis vectors, such that
\beq
    X^{(i)}_{a} \approx \sum_{\alpha = 1}^{N_\mathrm{SV}} c^{(i)}_\alpha V_{\alpha a}
\eeq
for $N_\mathrm{SV}< N_\mathrm{bin}$, where we assume that the SVs are arranged in decreasing order.\footnote{\new{By the Eckart-Young theorem, this is the \textit{best} rank-$N_\mathrm{SV}$ matrix approximation to $X$ \citep{eckart-young}.}} This is possible since $|c_\alpha|\leq D_\alpha$ as $U$ is unitary, and $D_\alpha$ is small for large $\alpha$. \resub{We henceforth drop the components of $U$, $V$ and $D$ with indices above $N_\mathrm{SV}$.} The projection corresponds to projecting the spectra into an $N_\mathrm{SV}$ dimensional Euclidean subspace, as we approximate
\beq\label{eq: inner-sv}
    \av{X^{(i)}|X^{(j)}} \approx \sum_{\alpha = 1}^{N_\mathrm{SV}} c_\alpha^{(i)}c_\alpha^{(j)}.
\eeq
\resub{Here, the first mode ($\alpha = 1$) defines the basis vectors which contribute most to the distance between two points on the manifold (or equivalently, the $\chi^2$ difference), the second mode is the transformation that has the second largest contribution \textit{et cetera.}}

We may fix $N_\mathrm{SV}$ by again considering the inner product. The mean squared distance of all $N_\mathrm{bank}$ spectra from the mean spectrum $\overline{P}$ is given by the average $\chi^2$;
\beq\label{eq: N-sv-def}
    \overline{\chi^2} &=& \frac{1}{N_\mathrm{bank}}\sum_i \av{\delta P^{(i)},\delta P^{(i)}}_\mathsf{C} = \frac{1}{N_\mathrm{bank}}\sum_{ia}X^{(i)}_aX^{(i)}_a\\\nonumber
    &=& \frac{1}{N_\mathrm{bank}}\sum_{ia\alpha\beta}U_{i\alpha}D_\alpha V_{\alpha a}U_{i\beta}D_\beta V_{\beta a}
    = \frac{1}{N_\mathrm{bank}}\sum_{i\alpha} U^2_{i\alpha}D_\alpha^2 = \frac{1}{N_\mathrm{bank}}\sum_\alpha D_\alpha^2
\eeq
since $V$ and $U$ are unitary. If one uniformly increases the number of template spectra across the manifold-with-boundary by a factor $f$, the mean $\chi^2$ (which is a geometric property of the manifold, unrelated to the template bank in the large $N_\mathrm{bank}$ limit) should be invariant, thus $D_\alpha \propto \sqrt{f}$. We thus conclude that $D_\alpha \propto \sqrt{N_\mathrm{bank}}$. Assuming $\overline{\chi^2}$ to be a good estimator of the distance between any two spectra in the template bank (and hence any two theory models on the prior-bounded manifold), the value of $N_\mathrm{SV}$ may be set by excluding those SVs which, in total, contribute less than some fixed $\chi^2_\mathrm{min}$ to $\overline{\chi^2}$; this fixes the optimal number of SVs, $N^*_\mathrm{SV}$, to
\beq\label{eq: N-sv-lim}
    N_\mathrm{SV}^* = \operatorname{min}\left\{N_\mathrm{SV} \left| \sum_{\alpha = N_\mathrm{SV}+1}^{N_\mathrm{bin}}D_\alpha^2 \leq \chi^2_\mathrm{min}N_\mathrm{bank}\right.\right\}.
\eeq
\resub{This condition ensures that the decomposition incurs a $\chi^2$-error below $\chi^2_\mathrm{min}$, averaged across the prior space.}
For a perfect measurement, with non-degenerate parameters, we would additionally require $N_\mathrm{SV}^* \geq N_\mathrm{param}$, where $N_\mathrm{param}$ is the dimensionality of the manifold. This would ensure that features of the spectra are not lost under the subspace projection, though, in practice, parameter degeneracies limit this.

A few comments on the SVD are needed. Firstly, we note that the decomposition of Eq.\,\ref{eq: SVD} implies
\beq
X^TX = V^TD^2V
\eeq
(keeping co-ordinates implicit for clarity), which is just a diagonalization with eigenvalue matrix $D^2$. From this, it is clear that the SVD performs an eigendecomposition (or equivalently, a Principal Component Analysis; PCA) of $X^TX$; the covariance matrix of (whitened) model spectra $X(\vec\theta)$ across the prior manifold.\footnote{\resub{This decomposition is not a new concept; such an approach forms the basis of many dimensionality reduction algorithms used in machine learning. For an additional application of this to approximating cosmological theory models, see Ref.\,\citep{2007ApJ...654....2F}.}} The first basis vector in $V$ (with largest singular value) therefore identifies the mode of the power spectrum that sources most of the variance between all templates sampled from the model space; the second vector identifies the second largest mode, subject to being orthogonal to the first one, and so on. Keeping only the basis vectors corresponding to the largest singular values (\textit{i.e.} Eq.\,\ref{eq: inner-sv}) thus encapsulates the main variation in $P(k)$ seen across the sampled model space. Given that we weight by $\mathsf{C}^{-1/2}$, this is equivalent to finding the modes which contribute most to the log-likelihood, $\chi^2$.

\new{Importantly, this approach is different from a standard eigendecomposition or PCA of the \emph{measurement} covariance matrix \citep[e.g.,][]{2000ApJ...544..597S,2013MNRAS.429..344K}, which identifies the orthogonal modes of the statistic that explain most of the variance of the measurement when sampling over different random realizations of the measurement rather than different samples from the model space. Unlike our approach, this PCA is agnostic of the model space, and thus cannot exploit the underlying structure, e.g., that cosmological parameters typically change multiple $k$-bins simultaneously in a smooth manner (implying that neighboring bins are correlated with respect to parameters, even on very large scales). Whilst the approach of covariance PCAs is to discard \textit{any} information corresponding to modes that are too noisy to be measured well, the effect of using SVDs is instead to discard any information which does not affect \textit{our} analysis, made possible by sampling only a finite region of theory space. Whilst this may lead to loss of information regarding unsampled parameters, it ensures that the basis vectors do not include much information about parameters for which we have strong prior knowledge}. \resub{Of course, our compression is optimal only for analyses that utilize the sampled parameters (or some transformations thereof); in practice, we expect the decomposition to still be useful in a larger parameter space, though the optimal analysis would require creation of a new template bank including the additional degrees of freedom.}

\new{An additional point of note concerns the choice of parameters, $\{\vec\theta^{(i)}\}$, with which we generate the template bank. A simple choice would be to draw samples uniformly from each cosmological parameter (subject to some limits), though there is freedom in their exact form, for example whether to use $\Omega_m$ or $\omega_{cdm}$, $A_s$ or $\log A_s$. In the general case, this will have non-trivial impact on the basis vectors, since it will give different weights to different sections of the manifold from which $X$ is sampled. However, we note that the SVD does not receive any information on the input parameters, only the output spectra, and thus, if the locations on the manifold are held fixed, the decomposition is invariant of the choice of co-ordinate chart. Furthermore, we expect identical results (on average) when drawing samples from any linear transform of the input parameters (assuming that the manifold boundary remains unchanged). In general however, we suggest using the same choices of parameters and priors (or at least a superset of these) to draw $\{\vec\theta^{(i)}\}$ as will be used in the later MCMC analysis, such that the basis vectors well represent the case in hand.} \resub{This blindness to the choice of parameters highlights another important note; the method is fully applicable to models which are non-linear in parameters, and thus give non-Gaussian posteriors. As an example, consider a parameter, $\zeta$, which enters the model quadratically. Based on the above discussion, if the sampling points on the prior manifold are identical, we can parametrize by either $\zeta^2$ or $\zeta$ and obtain the same decomposition. In the latter case, the posterior is highly non-Gaussian and indeed bimodal due to the $\zeta\leftrightarrow-\zeta$ symmetry.}

Whilst our decomposition is blind to the input parameters, there exist alternative approaches that include such information, for example partial least squares and canonical correlation analyses \citep{tenenhaus1998regression}. In these formalisms, both the input parameters and model data-vector are projected into a latent space; the benefit of this is that it gives the linear combinations of parameters that define the best-constrained features in the data-vector. Whilst this sounds appealing, in our context its interpretation is difficult since the optimal parameter sets are non-trivial combinations of cosmological and nuisance parameters, the latter of which are less meaningful.

\subsection{Application to Observational Likelihoods}\label{subsec: obs-likelihoods}
Whilst the above is somewhat abstract, it is of use when we consider observational data, since it gives us a rigorous method for which to reduce the dimensionality of the spectra. For a model power spectrum $\hat{P}$, the log-likelihood of parameters $\vec\theta$ given data covariance $\mathsf{C}_D$ is simply
\beq\label{eq: standard-log-like}
    -2\log\mathcal{L}(\vec\theta) = \hat\chi^2(\vec\theta) =  \sum_{ab}\left(\hat{P}_a-P_a(\vec\theta)\right)\mathsf{C}^{-1}_{D,ab}\left(\hat{P}_b-P_b(\vec\theta)\right),
\eeq
\resub{where we have assumed a Gaussian noise model, as is common in cosmology, and ignored the effects of parameter priors.} Under the previous rotation by fiducial covariance $\mathsf{C}$ (Eq.\,\ref{eq: X-def}), this can be written
\beq
    \hat\chi^2(\vec\theta) = \sum_{\alpha\beta}\left(\hat{X}_\alpha - X_\alpha(\vec\theta)\right)\left[\mathsf{C}^{T/2}\mathsf{C}^{-1}_{D}\mathsf{C}^{1/2}\right]_{\alpha\beta}\left(\hat{X}_\beta - X_\beta(\vec\theta)\right) .
\eeq
Note this is \textit{only} diagonal if the fiducial covariance matches that of the data. When performing inference from a small number of mocks, the data covariance is not well known, thus it is preferred to use a smooth model for $\mathsf{C}$ instead, giving a slightly non-Euclidean space. In terms of the basis vectors of Eq.\,\ref{eq: basis-decomposition}, and using modes only up to $N_\mathrm{SV}$, we obtain
\beq\label{eq: chi2-data}
    \hat\chi^2(\vec\theta) &\approx& \sum_{\alpha=1}^{N_\mathrm{SV}}\sum_{\beta=1}^{N_\mathrm{SV}}\left(\hat{c}_\alpha - c_\alpha(\vec\theta)\right)\left[V\mathsf{C}^{T/2}\mathsf{C}^{-1}_{D}\mathsf{C}^{1/2}V^T\right]_{\alpha\beta}\left(\hat{c}_\beta - c_\beta(\vec\theta)\right)\\\nonumber
    &=& \sum_{\alpha=1}^{N_\mathrm{SV}}\sum_{\beta=1}^{N_\mathrm{SV}}\left(\hat{c}_\alpha - c_\alpha(\vec\theta)\right)\mathcal{C}^{-1}_{D,\alpha\beta}\left(\hat{c}_\beta - c_\beta(\vec\theta)\right).
\eeq
In the second line, we have noted this likelihood is simply a Gaussian likelihood for the $\hat{c}$ variables, given covariance $\mathcal{C}_D$, which is simply the full covariance $\mathsf{C}_D$ projected into the subspace. For $\mathsf{C}_D = \mathsf{C}$, it is simply a unit matrix. Note that this can be written in terms of the inner product of Eq.\,\ref{eq: inner-product-X} as
\beq\label{eq: chi2-data2}
    \hat\chi^2(\theta) = \av{\hat{X}-X(\vec\theta)|\hat{X}-X(\vec\theta)}_{D}
\eeq
where the subscript $D$ indicates that this is with respect to the metric $g_{\alpha\beta} = \mathcal{C}^{-1}_{D,\alpha\beta}$, rather than being strictly Euclidean.

Given a set of mocks, we may estimate the projected covariance directly from the measurements of $\hat{c}$ in each.\footnote{It is worth noting that the subspace coefficients, $c$, do not have straightforward correspondences with individual cosmological parameters. As an example, consider an analysis measuring only the primordial amplitude. As required by the SVD, the first coefficent dominates, but will have contributions from \textit{all} nuisance parameters that affect the amplitude, especially those that control the high-$k$ regime, where the error bars are small.} Since this contains fewer bins than the unprojected covariance, it is less susceptible to stochastic shifts arising from covariance matrix noise. The inference proceeds by computing the model power spectrum for each chosen parameter vector, projecting onto the reduced $V_\alpha$ basis, and computing the likelihood of Eq.\,\ref{eq: chi2-data} using the mock-based $\mathcal{C}_D$ subspace covariance. MCMC can then be performed using this subspace likelihood.\footnote{Note that it is not sufficient to simply use the template bank samples to compute the likelihood posterior surface (as in done in gravitational wave analyses, e.g., Ref.\,\citep{2019PhRvD..99l3022R}). For a posterior that is compact in the prior space, there are very few template bank samples near the minimum point of the likelihood. In this example, we have a minimum log-likelihood of 300 in the data, versus just three in the MCMC output. This could be reduced by using more restrictive priors on the template bank.} Subject to a suitably chosen $N_\mathrm{SV}$, we expect the incurred $\chi^2$ error to be small, and thus the posteriors to be unbiased.

\section{Parameter Shifts and Covariances}\label{sec: shifts}
\subsection{Noise-Averaged Constraints}\label{subsec: theory-no-noise}
It is important to show that the $\vec\theta$ estimates obtained in this subspace formalism are unbiased, and produce a good estimate of the parameter covariance. To do this, we write the data coefficients as $\hat{c} = c(\vec\theta^*)+\hat{n}$ where $\theta^*$ are the true parameters and $\hat{n}$ is some noise vector, assumed to be uncorrelated with theory. For the simple case of matching true and fiducial $P(k)$ covariances ($\mathsf{C}=\mathsf{C}_D$), we have a diagonal subspace metric $\mathcal{C}^{-1}_{D,\alpha\beta} = \delta^K_{\alpha\beta}$, which implies that each element of $\hat{n}$ is independent and drawn from a Gaussian of unit variance. In contrast, due to the use of SVD to define the subspace basis vectors, we expect the magnitude of the $c_\alpha(\theta^*)$ coefficients to fall strongly with $\alpha$; we thus expect the signal-to-noise of the modes to fall monotonically, such that the later components are noise dominated and can be justifiably removed.

The subspace log-likelihood is given by Eq.\,\ref{eq: chi2-data}, which we may write as another inner product in the space spanned by $c(\vec\theta)$;
\beq\label{eq: chi2cform}
    \hat\chi^2(\vec\theta) &=& \inn{\hat{c}-c(\vec\theta)}{\hat{c}-c(\vec\theta)}\\\nonumber
    &=& \inn{c(\vec\theta^*)-c(\vec\theta)+\hat{n}}{c(\vec\theta^*)-c(\vec\theta)+\hat{n}}.
\eeq
The observed mean parameter vector $\hat\theta$ is defined by maximizing $\hat\chi^2$;
\beq    
    \left.\frac{\partial\chi^2}{\partial\theta_i}\right|_{\vec\theta=\hat{\vec\theta}} = 0 \quad \Rightarrow \quad \left.\inn{\frac{\partial c(\vec\theta)}{\partial\theta_i}}{c(\vec\theta^*)-c(\hat{\vec\theta})-\hat{n}}\right|_{\vec\theta=\hat{\vec\theta}} = 0,
\eeq
where $i\in \{1,...,N_\mathrm{param}\}$ indexes the parameter vector.\footnote{\resub{Throughout this section we assume that any Gaussian parameter priors (e.g., those on the higher-order bias parameters) are not strongly informative and may be ignored; a justifiable assumption in this work.}} Averaging over noise, this simply requires $c(\vec\theta^*)-c(\hat{\vec\theta}) = \vec0$, implying that $\hat{\vec\theta} = \vec\theta^*$ (assuming that the mapping $\vec\theta\rightarrow c(\vec\theta)$ is \new{injective, which is usually the case in cosmological contexts}).
Note this is true for any value of $N_\mathrm{SV}$; \textit{i.e.} the subspace decomposition always gives an unbiased estimate of the parameters, as required.

For the parameter covariance, we start from the Fisher matrix, defined by
\beq
    2\hat{\mathcal{F}}_{ij} = \left.\frac{\partial^2\hat{\chi}^2(\vec\theta)}{\partial\theta_i\partial\theta_j}\right|_{\vec\theta=\hat{\vec\theta}} = \inn{\frac{\partial c(\hat{\vec\theta})}{\partial \theta_i}}{\frac{\partial c(\hat{\vec\theta})}{\partial \theta_j}} + \inn{\frac{\partial^2 c(\hat{\vec\theta})}{\partial \theta_i\partial\theta_j}}{c(\vec\theta^*)-c(\hat{\vec\theta})-\hat{n}} + (i\leftrightarrow j).
\eeq
Averaging over noise allows us to drop the second set of parentheses (since $\hat{\vec\theta} = \vec\theta^*$). From this, we can see that using less than $N_\mathrm{bin}$ SVs (\textit{i.e.} restricting to a subspace) produces a shift in the noise-averaged Fisher matrix;
\beq
    \Delta\mathcal{F}_{ij} =  \left(\sum_{\alpha=1}^{N_\mathrm{SV}}\sum_{\beta=1}^{N_\mathrm{SV}} - \sum_{\alpha = 1}^{N_\mathrm{bin}}\sum_{\beta = 1}^{N_\mathrm{bin}}\right)\frac{\partial c_\alpha(\vec\theta^*)}{\partial \theta_i}\mathcal{C}_{D,\alpha\beta}^{-1}\frac{\partial c_\beta(\vec\theta^*)}{\partial \theta_j}.
\eeq
This generates a corresponding shift in the noise-averaged parameter covariance $\Phi = \mathcal{F}^{-1}$;
\beq\label{eq: shift-parameter-cov}
    \Delta\Phi_{ij} = -\sum_{kl}\mathcal{F}^{-1}_{ik}\Delta\mathcal{F}^{ }_{kl}\mathcal{F}^{-1}_{lj} = -\sum_{kl}\Phi_{ik}\Delta\mathcal{F}_{kl}\Phi_{lj}.
\eeq
As shown in Appendix \ref{appen: fisher-matrices}, $\Delta\Phi_{ij}$ is positive semidefinite; \textit{i.e.} restricting to a non-trivial subspace can only increase the parameter covariance. Whilst this shift exists, it is expected to be small, since the $c_\alpha$ coefficients rapidly decrease with index $\alpha$, thus we expect similar behavior for the parameter derivatives and hence contributions to $\Delta \Phi_{ij}$.

\subsection{Noise in the Data Vector}\label{subsec: theory-noisy-data}
Whilst the above subsection gives the behavior of the subspace likelihood averaged over noisy realizations, it remains to consider how the best-fit parameter is shifted from the mean for individual noise realizations. Assuming the inverse covariance matrix $\mathcal{C}_D$ to be known precisely, Eq.\,\ref{eq: parameter-error} of Appendix \ref{appen: param-shifts} shows the shift in best-fit parameter for a given noise realization $\hat{n}$ to be equal to
\beq\label{eq: param-shift-no-cov}
    \delta\theta_i = \sum_{j}\inn{\frac{\partial c(\vec\theta^*)}{\partial\theta_j}}{\hat{n}}\Phi_{ij}
\eeq
adopting the inner product of Eq.\,\ref{eq: chi2cform} and neglecting (subdominant) noise in the parameter covariance, $\Phi$. Since this is a stochastic quantity it is best understood by computing the shift covariance, $\av{\delta\theta_i\delta\theta_j}$. 
Using the definition $\av{\hat{n}_\alpha\hat{n}_\beta} = \mathcal{C}_{D,\alpha\beta}$, the average can be simply obtained;
\beq
    \av{\delta\theta_i\delta\theta_j} = \Phi_{ij}.
\eeq
As noted in the previous section, $\Phi_{ij}$ increases as $N_\mathrm{SV}$ is decreased, thus a reduction in the number of basis vectors necessarily increases the magnitude of best-fit parameter shifts. However, given that $c_\alpha(\theta^*)$ falls rapidly with increasing index $\alpha$ (due to the SVD), we expect higher SVs to have only small contributions to the shift covariance.

\subsection{Noise in the Precision Matrix}\label{subsec: theory-noisy-precision}
Of greater importance in this work is noise in the precision matrix, which generally appears when estimating $\mathcal{C}_D^{-1}$ using a finite number, $N_\mathrm{mock}$, of mocks. As shown in Eq.\,\ref{eq: cov-noise-shift} in Appendix \ref{appen: param-shifts}, a \resub{stochastic shift} in the precision matrix, $\delta\Psi$, leads to an additional shift in the best-fit parameters
\beq\label{eq: best-fit-shift}
    \Delta\theta_i = \sum_{\alpha\beta j}\left[\Phi_{ij}\hat{n}_\beta\frac{\partial c_\alpha}{\partial\theta_j} - \sum_{kl} \Phi_{ik}\Phi_{jl}\inn{\frac{\partial c}{\partial\theta_j}}{\hat{n}}\frac{\partial c_\alpha}{\partial\theta_k}\frac{\partial c_\beta}{\partial\theta_l}\right]\delta\Psi_{\alpha\beta},
\eeq
\resub{valid for any likelihood monotonic in $\hat{\chi}^2$.} Whilst the exact form is unimportant for our purposes here, we note that it depends on the product $\delta\Psi\hat{n}$; there is no shift from noisy precision matrices if the data is noiseless. For noisy data, this shift is present, though it vanishes on average providing that $\av{\delta\Psi} = \av{\delta\Psi \hat{n}} = 0$.\footnote{These conditions are non-trivial in some cases. Systematic parameter biases can be obtained from smooth models of the covariance matrix which have $\av{\delta\Psi}\neq 0$ as well as covariance matrices estimated using the data, which have $\av{\delta\Psi\hat{n}}\neq 0$.} Averaging over statistical noise in both $\hat{n}$ and $\delta\Psi$ (assuming Gaussian statistics), we obtain the shift covariance from precision matrix noise;
\beq\label{eq: shift-covariance-factor}
    \av{\Delta\theta_i\Delta\theta_j} = \frac{(N_\mathrm{mock}-N_\mathrm{SV})(N_\mathrm{SV}-N_\mathrm{param})}{(N_\mathrm{mock}-N_\mathrm{SV}-1)(N_\mathrm{mock}-N_\mathrm{SV}-4)}\Phi_{ij}
\eeq
\citep{2013PhRvD..88f3537D,2013MNRAS.432.1928T}, which includes the Hartlap factor required to debias the inversion of noisy covariance matrices \citep{2007A&A...464..399H}. In the limit of large $N_\mathrm{mocks}\gg N_\mathrm{SV}$, this is approximately
\beq\label{eq: rough-inflation-factor}
    \av{\Delta\theta_i\Delta\theta_j} \approx \frac{N_\mathrm{SV}-N_\mathrm{param}}{N_\mathrm{mock}}\Phi_{ij}.
\eeq
It is here that we see the utility of using $N_\mathrm{SV}<N_\mathrm{bin}$. This dramatically decreases the parameter \resub{shifts} obtained using small $N_\mathrm{mock}$, and, conversely, allows one to perform equally accurate inference using many fewer mocks, greatly improving the computational efficiency for surveys with a large number of bins. The standard way of accounting for such errors is to inflate the output parameter covariances \citep{2014MNRAS.439.2531P} and thus lose constraining power; using fewer SVs reduces this need, implying that the parameter confidence contours will \textit{reduce} as $N_\mathrm{SV}$ decreases.

\resub{The assumption of Gaussianity in the above discussion is not fully valid. As shown in Ref.\,\citep{2016MNRAS.456L.132S}, when the covariance matrix is estimated from a finite number of mocks, one should properly marginalize over its inverse, which leads to the likelihood being replaced by a multivariate t-distribution of the following form;
\beq\label{eq: like-sellentin-heavens}
    -2\log\mathcal{L}(\vec\theta)\rightarrow N_\mathrm{mock}\log\left(1+\frac{\hat{\chi}^2(\vec\theta)}{N_\mathrm{mock}-1}\right) + \text{const.}
\eeq
with no Hartlap factors required. In general, this correction becomes important when $N_\mathrm{mock}$ becomes comparable to $N_\mathrm{SV}$, and alters the tails of the posterior distributions, causing a significant inflation relative to that of the na\"ive Gaussian likelihood (without the Hartlap correction factor). Whilst Eq.\,\ref{eq: like-sellentin-heavens} is simple to implement in a sampling code, we principally adopt the Gaussian likelihood in this work, since the alternative complicates some technical points pertaining to the analytic marginalization over nuisance parameters discussed in Appendix \ref{appen: an-marg}. In fact this makes very little difference to the output parameter constraints; assuming uninformative priors, the likelihood of Eq.\,\ref{eq: like-sellentin-heavens} leads to the posterior distribution itself being a multivariate t-distribution with $N_\mathrm{mock}-N_\mathrm{param}$ degrees of freedom. In general, this is large, hence the posteriors are very close to Gaussian, and we find very similar parameter constraints from the Hartlap-rescaled Gaussian likelihood and that of Ref.\,\citep{2016MNRAS.456L.132S}. Furthermore, in both choices of likelihood we find stochastic shifts in the \textit{best-fit} parameters arising from covariance matrix noise (Eq.\,\ref{eq: best-fit-shift}). Several works advocate for the inclusion of an additional \textit{inflation factor}, $m_1$, to ensure that the output parameter variances match those one would obtain by repeating the analysis many times allowing for stochastic noise in both the data-vector and covariance. Here, we adopt the rescaling factor of Ref.\,\citep{2014MNRAS.439.2531P}, which is derived in the Gaussian limit and discussed further in Sec.\,\ref{subsec: data-prec-noise}.}

\resub{We emphasize that such posterior inflation is not strictly mathematically justified. Indeed, the inflation factor, $m_1$, essentially captures the the difference between the 
true parameter variance obtained with a correct covariance 
matrix (\textit{i.e.} one from infinitely many mocks) and that averaged over a large number of covariance matrix realizations. However, neither a set of such realisations nor the true covariance are available in actual 
parameter estimates. $m_1$ should thus be viewed as a phenomenological factor introduced in order to 
approximately compensate for the stochastic shifts of the best-fit by the sampling noise in the covariance matrix. Whilst this is the approach adopted by the BOSS and eBOSS collaborations, it is necessarily inferior to a fully Bayesian treatment.}

\subsection{Choice of Fiducial Covariance Matrix}\label{subsec: theory-fiducial-choice}
The choice of fiducial covariance $\mathsf{C}$ used to generate the SVD subspace warrants discussion. For $N_\mathrm{SV} = N_\mathrm{bin}$, this is arbitrary since the mapping from power spectra to coefficient space is invertible; equivalently, $\chi^2$ is independent of $\mathsf{C}$. In the general case it is important, since the fiducial covariance defines the mapping onto the lower-dimensional subspace. If $\mathsf{C}$ is equal to the power spectrum data covariance $\mathsf{C}_D$, the subspace metric is Euclidean and all $c_\alpha$ coefficients are uncorrelated. Since we define basis vectors via SVD, we expect the information content of the basis coefficients to decrease monotonically with the index $\alpha$; having a diagonal covariance ensures that we do not throw away information arising from correlations between low and high basis coefficients. In general, the subspace metric is defined by the projection
\beq\label{eq: cov-projection}
    \mathcal{C}_{D}^{-1} = V\mathsf{C}^{T/2}\mathsf{C}^{-1}_D\mathsf{C}^{1/2}V^T
\eeq
(Eq.\,\ref{eq: chi2-data}), which becomes more diagonal as $\mathsf{C}$ approaches $\mathsf{C}_D$ (since $V$ is unitary). For this reason, using a poor estimate of $\mathsf{C}_D$ will lead to significant correlations between basis coefficients, and thus reduce the efficacy of the method by introducing a larger shift in the parameter covariance (Eq.\,\ref{eq: shift-parameter-cov}) as SVs are removed. It is thus desirable to use some input estimate of $\mathsf{C}$ which is close to the truth (to allow for efficient subspace projections) and low-noise (to ensure the basis vectors are smooth). It is worth noting however, that \textit{any} invertible choice of fiducial covariance gives unbiased parameter estimates in the limit of zero noise; optimizing this simply ensures that the posteriors are less affected by noise for a given (small) number of basis vectors.


\section{Application to BOSS DR12}\label{sec: applications}
\subsection{Mock Data-sets and Covariances}
We apply the formalism of the above section to galaxy power spectra taken from the twelfth data release (DR12) \citep{2017MNRAS.470.2617A} of the Baryon Oscillation Spectroscopic Survey (BOSS), which is part of SDSS-III \citep{2013AJ....145...10D,2011AJ....142...72E}. For simplicity, we use only the largest of the four data-chunks, the high-$z$ North Galactic Cap sample, with mean redshift $z = 0.61$ and volume $V = 2.8h^{-3}\mathrm{Gpc}^3$. In this work, all mock data is drawn from the publicly available\footnote{\href{https://fbeutler.github.io/hub/boss_papers.html}{https://fbeutler.github.io/hub/boss\_papers.html}} power spectrum multipoles from 2048 MultiDark-Patchy mocks (hereafter Patchy) \citep{2014MNRAS.439L..21K,2016MNRAS.456.4156K,2016MNRAS.460.1173R}, including 48 $k$-bins in $[0.01,0.25]\hMpc$ for both monopole and quadrupole moments, as in Refs.\,\citep{2020JCAP...05..042I,2020PhRvD.101h3504I,2020JCAP...05..032P,2020arXiv200808084P}. Two choices of mock data are used: a single Patchy realization, which emulates the BOSS sample, and the mean-of-48 realizations, to ensure our analysis is unbiased.

For the covariance matrix, we consider three possible choices: (1) the sample covariance from Patchy mocks; (2) the analytic covariance presented in Ref.\,\citep{2019arXiv191002914W}; and (3) a simplified Gaussian (plus shot-noise) covariance computed for the best-fit BOSS cosmology. In the former case, we use either $N_\mathrm{mock} = 2000$ or $125$ mocks (excluding those used to define the data-vectors), defining the \resub{sample} covariance via the usual formula
\beq\label{eq: cov-Pk}
    \operatorname{cov}\left(P_a,P_b\right) \equiv \mathsf{C}_{D,ab} = \frac{1}{N_\mathrm{mock}-1}\sum_{i = 1}^{N_\mathrm{mock}}\left(\hat{P}^{(i)}_a - \overline{P}_a\right)\left(\hat{P}^{(i)}_b - \overline{P}_b\right),
\eeq
where $\hat{P}^{(i)}_a$ is the power spectrum of the $i$-th mock, subscripts indicate the $k$ bin and multipole, and the overbar represents the mean over all mocks. When forming $\chi^2$ we require the inverse covariance, $\mathsf{\Psi}_D$; to ameliorate noise-induced bias, we include the Hartlap factor \citep{2007A&A...464..399H}, defined by
\beq\label{eq: hartlap}
    \mathsf{\Psi}_D = f_\mathrm{H}\times \mathsf{C}_D^{-1},\quad f_\mathrm{H} = \frac{N_\mathrm{mock}-N_\mathrm{bin}-2}{N_\mathrm{mock}-1}.
\eeq
Since the other choices of covariance are analytic, they do not require a Hartlap factor. The first of these uses perturbation theory to compute the off-diagonal trispectrum terms and the contributions from super-survey modes, whilst the diagonal spectrum is estimated from Patchy. In contrast, the second assumes Gaussianity (and thus no off-diagonal terms before window convolution), with the diagonal computed via the one-loop theory model used in this work. Both include the BOSS window function, as discussed in Ref.\,\citep{2019arXiv191002914W}.

\subsection{Creation of the Template Bank and Subspace Coefficients}\label{sec: template-creation}
The template bank is generated following the procedure of Sec.\,\ref{subsec: template-bank}. We first draw $10^5$ points in the parameter space of Eq.\,\ref{eq: parameter-space}, subject to the broad priors;
\beq\label{eq: param-priors}
    &&h\in [0.6,0.74],\quad \omega_{cdm}\in [0.08,0.16], \quad A_s/A_{s,\mathrm{fid}}\in [0.6,1.4], \quad b_1\in [1.7,2.3],\\\nonumber
    &&b_2\sim \mathcal{N}(0,1), \quad b_{G_2}\sim \mathcal{N}(0,1), \quad b_4\sim\mathcal{N}(500,500^2),\\\nonumber
    &&c_{s,0}\sim \mathcal{N}(0,30^2), \quad c_{s,2}\sim \mathcal{N}(0,30^2), \quad P_\mathrm{shot}\sim \mathcal{N}(0,5000^2),
\eeq
where $\mathcal{N}(\mu,\sigma^2)$ indicates a Gaussian prior and all quantities are in $\hMpc$-type units. Note that the flat priors are only for the generation of the template bank, and are not used in the later \new{MCMC}
analysis.\footnote{$10^5$ samples is likely overkill; the change to the dominant basis vectors is negligible when this is reduced to $10^4$ samples, \resub{and remains small even for $N_\mathrm{bank} = 10^3$. Both of these cases require the same number of basis vectors (12) to incur a prior-averaged $\chi^2$ error below $0.1$.}} For each set of parameters, the power spectrum model is computed via one-loop Effective Field Theory, as implemented in the \texttt{CLASS-PT} code \citep{2020arXiv200410607C}, then convolved with the BOSS window function. Note that the priors are purposefully very broad; given more restrictive priors, the domain of the parameter manifold can be reduced, leading to fewer required subspace coefficients. 

To compute the basis vectors of the subspace discussed in Sec.\,\ref{subsec: template-bank}, we first require the fiducial covariance $\mathsf{C}$, which defines the rotated power spectra $X$ (Eq.\,\ref{eq: X-def}). Whilst setting this equal to the sample covariance would ensure that the subspace metric is diagonal, this is seldom desirable since, unless $N_\mathrm{mock}$ is very large, it will lead to noisy basis vectors and hence less optimal subspace decompositions. In general, it is important to use a fiducial covariance that is (a) relatively smooth, and (b) provides a fair approximation to the true covariance. For this reason, we principally set $\mathsf{C}$ equal to the analytic covariance matrix of Ref.\,\citep{2019arXiv191002914W}, as described above.\footnote{Note that having $\mathsf{C}\neq \mathsf{C}_D$ does not bias our inference, though it changes the efficiency of our subspace projection. If we use the same number of SVs as bins, the likelihood is independent of our choice of $\mathsf{C}$. Assuming the sample covariance to be fixed, our analysis is robust to inaccuracies in the fiducial covariance.} 

Using this, we rotate the template spectra into the $X$ variables and apply SVD as in Eq.\,\ref{eq: SVD} to compute the set of basis vectors $\{V_\alpha\}$. This further defines the SVs $\{D_\alpha\}$ which can be used to set $N_\mathrm{SV}$.\footnote{\new{See Fig.\,\ref{fig: pkbk-svs} for a representative plot of the $D_\alpha$ components.}} \resub{To set $N_\mathrm{SV}$ we use Eq.\,\ref{eq: N-sv-def}, enforcing that $\chi^2$ be reproduced to within 0.1, averaged across the prior, which we expect to give reasonable results.} This requires $N_\mathrm{SV} \approx 12$ (depending on the exact covariance),; \textit{i.e.} all SVs above this index contribute less than 0.1 to the mean $\chi^2$. This corresponds to compressing the data by a factor of eight, and is slightly larger than the number of parameters (10), indicating the effects of non-linearities in the \new{transformation $\vec\theta\rightarrow P(\vec\theta)$ (and hence the likelihood) that cannot be well represented as linear across the prior domain. Alternatively, this signifies a break-down of the assumption that all information can be encapsulated in the rank-$N_\mathrm{param}$ Fisher matrix.}
Below, we will also consider $N_\mathrm{SV} = 48$ and the uncompressed power spectrum vector for testing purposes. Had we used more cosmological or nuisance parameters in our model, we would expect $N_\mathrm{SV}$ to increase correspondingly.

Given the above Patchy mocks, we generate corresponding subspace coefficients $\{\hat{c}_\alpha^{(i)}\}$ using the first $N_\mathrm{SV}$ basis vectors via Eq.\,\ref{eq: basis-decomposition}. These are then used to compute the sample covariance of the coefficients via the standard formula;
\beq
    \operatorname{cov}\left(c_\alpha,c_\beta\right) \equiv \mathcal{C}_{D,\alpha\beta} &=& \frac{1}{N_\mathrm{mock}-1}\sum_{i = 1}^{N_\mathrm{mock}}\left(\hat{c}^{(i)}_\alpha-\overline{c}_\alpha\right)\left(\hat{c}^{(i)}_\beta-\overline{c}_\beta\right)
\eeq
(cf.\,Eq.\,\ref{eq: cov-Pk}), which is used in the subspace likelihood (Eq.\,\ref{eq: chi2-data}). As for the former covariance, this requires a Hartlap factor to invert; the upside is that the Hartlap factor will be closer to unity here if $N_\mathrm{SV}<N_\mathrm{bin}$, shrinking the parameter covariances. Additionally, we will find it useful to perform the analysis using analytic covariances within the likelihood (rather than just as the fiducial covariance); in this case, we can form the subspace data covariance from Eq.\,\ref{eq: cov-projection}.

\subsection{Posterior Sampling Methods}
With the datasets and basis vectors in hand, posterior surfaces are computed using Markov Chain Monte Carlo (MCMC) methods, here implemented via \texttt{montepython} v3.3 \citep{2013JCAP...02..001A,2018arXiv180407261B}. Given the necessity to run MCMC a significant number of times to validate our method, we implement two approaches to expedite the sampling. Firstly, we note that any parameter which enters the power spectrum model $P(\vec\theta)$ linearly can be analytically marginalized without loss of information \citep{2010MNRAS.408..865T,2002MNRAS.335.1193B}. This procedure is described in Appendix \ref{appen: an-marg}, and simply leads to those parameters being absorbed into a cosmology-dependent covariance matrix. In our case, the nuisance parameters $b_4,c_{s,0},c_{s,2}$ and $P_\mathrm{shot}$ fall in this category, 
and are thus marginalized over directly. Note that this does not affect the SVD decompositions since it is applied only in the final MCMC step.

Secondly, when computing multiple posteriors at similar locations, it is illogical to rerun the MCMC each time from scratch. Instead, we opt to run the MCMC once, saving the (unprojected) power spectra for each point in parameter space, then use these samples to reconstruct the posterior for a different analysis, e.g., a different choice of $N_\mathrm{SV}$, via standard importance sampling techniques. This is fast (since it does not require a Boltzmann code to be run) and highly accurate for posteriors close to the initial MCMC chain. The latter requirement can be assessed by the \textit{effective sample size} of the resampled chain \citep[e.g.][]{2016arXiv160203572M}, equal to
\beq\label{eq: effective-sample-size}
    \mathrm{ESS} = \frac{\left(\sum_i w_i\right)^2}{\sum_i w_i^2},
\eeq
where $w_i$ are the importance sampling weights. For a new posterior, if the effective sample size is too small ($\lesssim 10^4$ samples) when importance sampling from any pre-existing set of spectral samples, we simply create a new MCMC chain for this configuration.

\subsection{Results}
Below, we present the results of the subspace analyses of the mock power spectrum data. Unless otherwise specified, we use a single realization of mock data, set the fiducial covariance to that of Ref.\,\citep{2019arXiv191002914W} (including both Gaussian and non-Gaussian components) and use a sample covariance from 2000 Patchy mocks. In all cases, we plot the derived parameters $H_0$, $\Omega_m$ and $\sigma_8$; in the true Patchy cosmology, these are given by $67.77$, $0.3071$ and $0.8288$ respectively.

\subsubsection{Bias in the Noiseless Limit}

\begin{figure}
\centering
\begin{minipage}[t]{.48\textwidth}
  \centering
  \includegraphics[width=\textwidth]{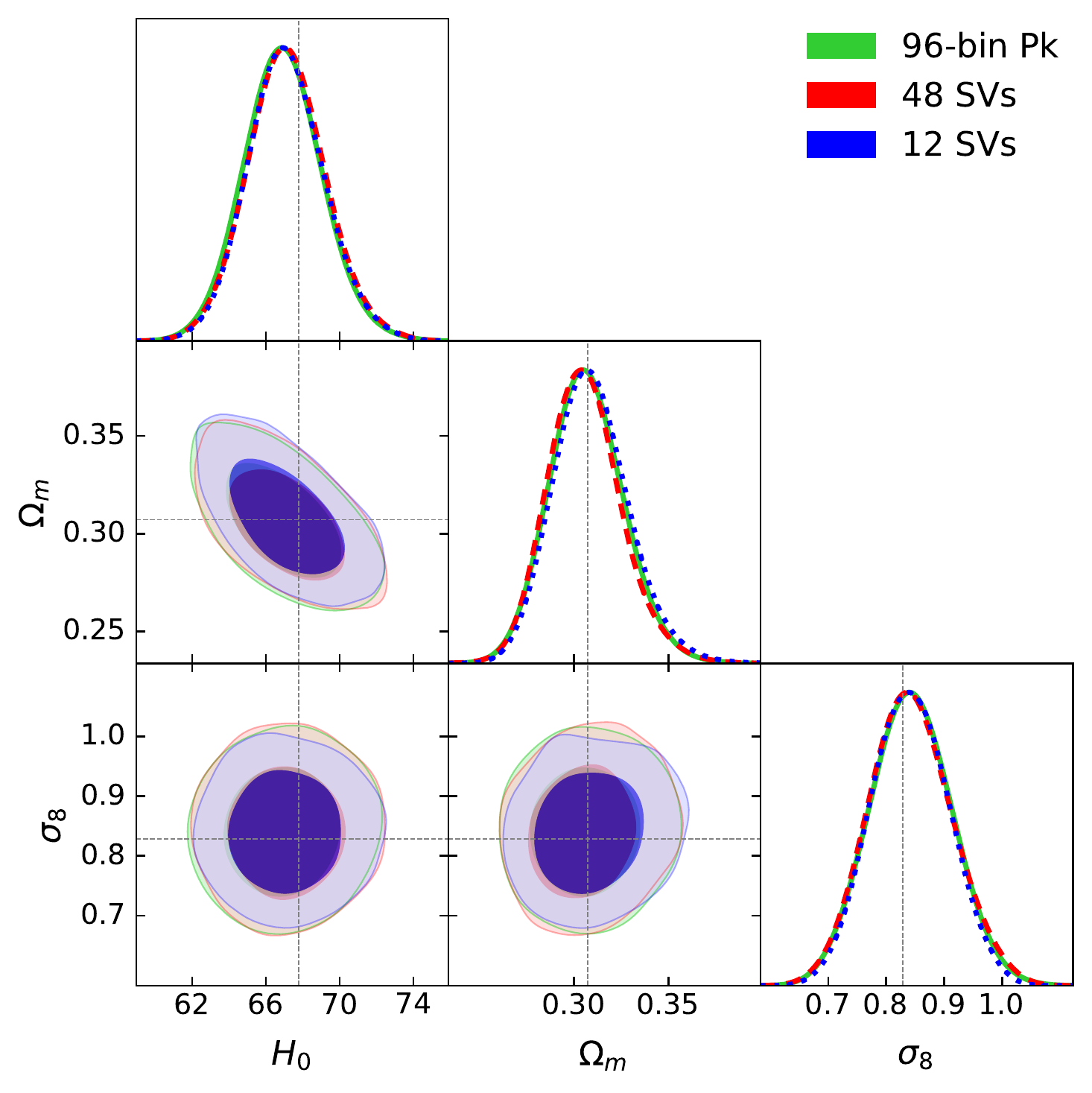}
  \caption{Corner plot of the MCMC posterior from analyzing the mean of 48 Patchy mocks, using both the standard analysis in 96 power spectrum bins (green, full lines), 48 subspace coefficients (red, dashed lines) and 12 subspace coefficients (blue, dotted lines). The true Patchy cosmology is indicated by the vertical lines in the 1D histograms. All plots are computed using a 2000-mock Patchy covariance matrix, with an analytic covariance used to define the SVD subspace projection, following the method discussed in Sec.\,\ref{sec: methodology}. The posterior is clearly unbiased in all cases.}
  \label{fig: mean-of-mocks}
\end{minipage}%
\quad
\begin{minipage}[t]{.48\textwidth}
  \centering
  \includegraphics[width=\textwidth]{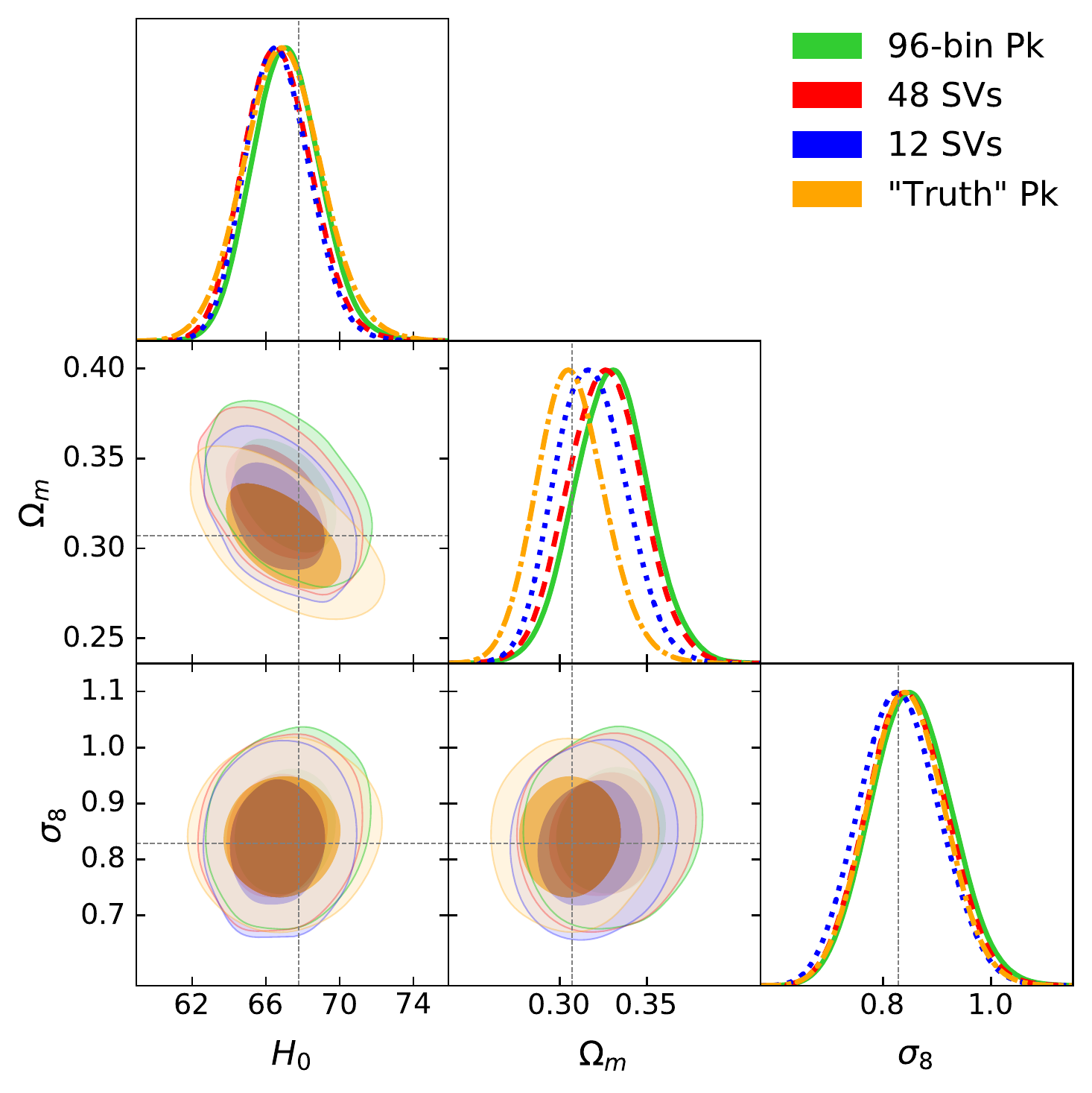}
  \caption{As Fig.\,\ref{fig: mean-of-mocks}, but analyzing a single Patchy mock. We additionally show the `true' results from the mean of 48 mocks in yellow (dot-dashed lines). Here, restricting to fewer basis coefficients does induce a shift in the parameters (as discussed in Sec.\,\ref{subsec: theory-noisy-data}), \new{due to slightly different noise properties, since fewer noisy modes are included.}
  }
  \label{fig: single-mock}
\end{minipage}
\end{figure}

The first important check is whether the subspace analysis gives an unbiased estimate of cosmological parameters in the limit of zero noise \resub{in the data-vector}. Fig.\,\ref{fig: mean-of-mocks} displays the posterior contours when the above analysis is applied to the mean-of-mocks data-set, comparing $96$ uncompressed power spectrum bins, $48$ SVs and $12$ SVs. Comparing the modal parameters to the true Patchy cosmology (from which the mock data is generated) shows excellent agreement in all cases; furthermore the posterior shapes are consistent between all data-sets, with no noticeable increase in the parameter variances due to the subspace projections. This matches the theoretical calculation of Sec.\,\ref{subsec: theory-no-noise}, which asserted that the subspace likelihood would give an unbiased estimate of the mean, but a small increase to the parameter covariance; we here confirm that this increase is negligibly small even when using $N_\mathrm{SV} = 12$.

\subsubsection{Bias from a Single Realization}
Next, we consider the analogous results using data from a single Patchy mock, emulating the true observational sample. \resub{This uses the 2000-mock sample covariance, thus we expect the effects of parameter shifts induced by covariance matrix noise to be small.} As discussed in Sec.\,\ref{subsec: theory-noisy-data}, we expect some stochastic parameter shifts as a result of the differing noise properties, since data-vectors with lower $N_\mathrm{SV}$ include fewer noise-dominated modes. 
These results are shown in Fig.\,\ref{fig: single-mock} and confirm our suspicions; there is a noticeable shift in parameters (in particular $\Omega_m$) as we move from 96 to 48 to 12 bins in the data-vector. \resub{Whilst in this example the posteriors shift in the direction of the noiseless limit, we expect the sign of this shift to vary for different noise realizations.}
In any case, although the noise properties of the subspace likelihoods differ somewhat from those of the true data, the results of the previous subsection confirm to us that the analysis is not biased as a result.

\subsubsection{\resub{Parameter Shifts} from Precision Matrix Noise}\label{subsec: data-prec-noise}

\begin{figure}%
    \centering
    \subfloat[96-bin Power Spectrum]{{\includegraphics[width=0.3\textwidth]{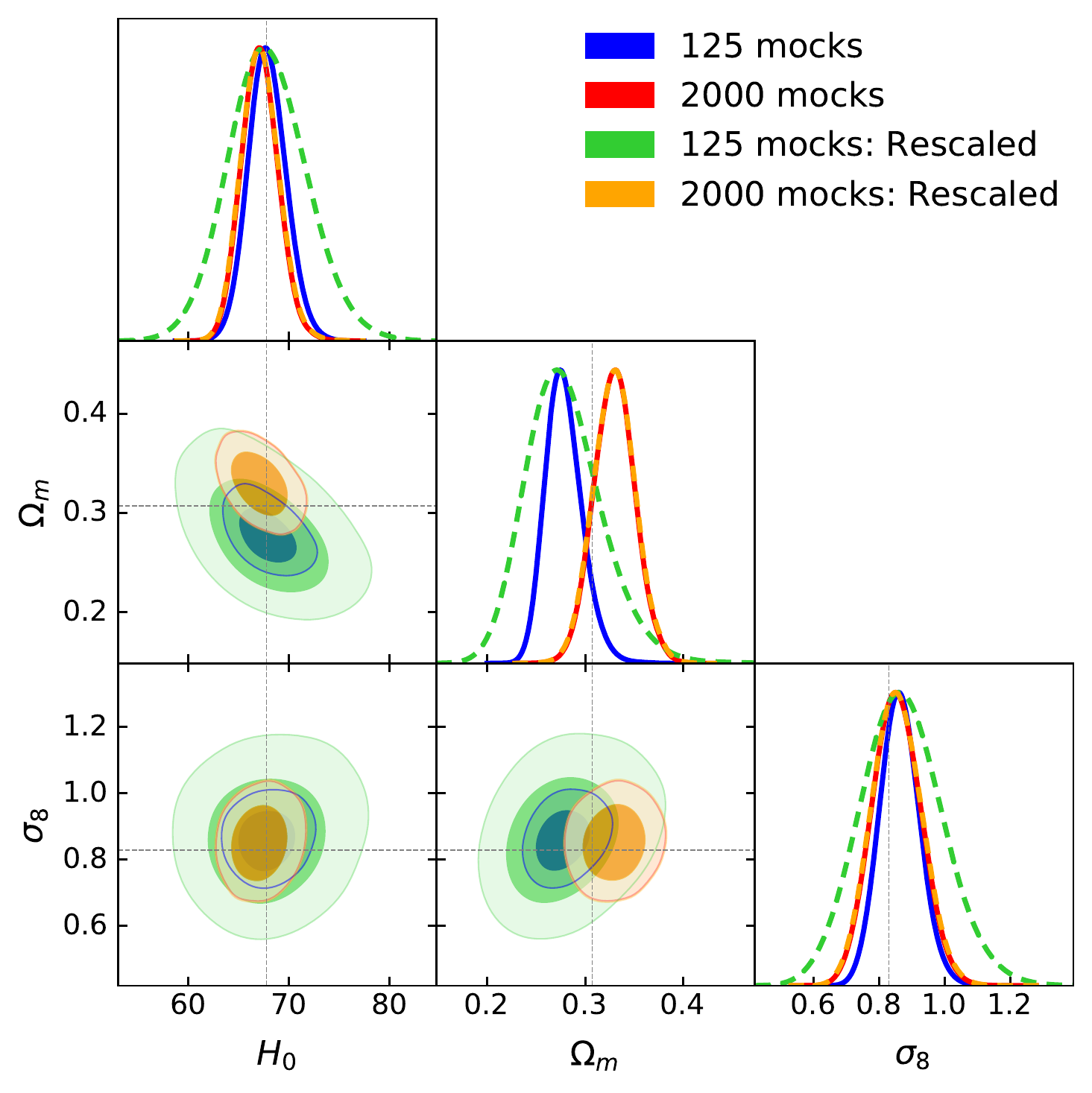} }}%
    \qquad
    \subfloat[48 Subspace Coefficients]{{\includegraphics[width=0.3\textwidth]{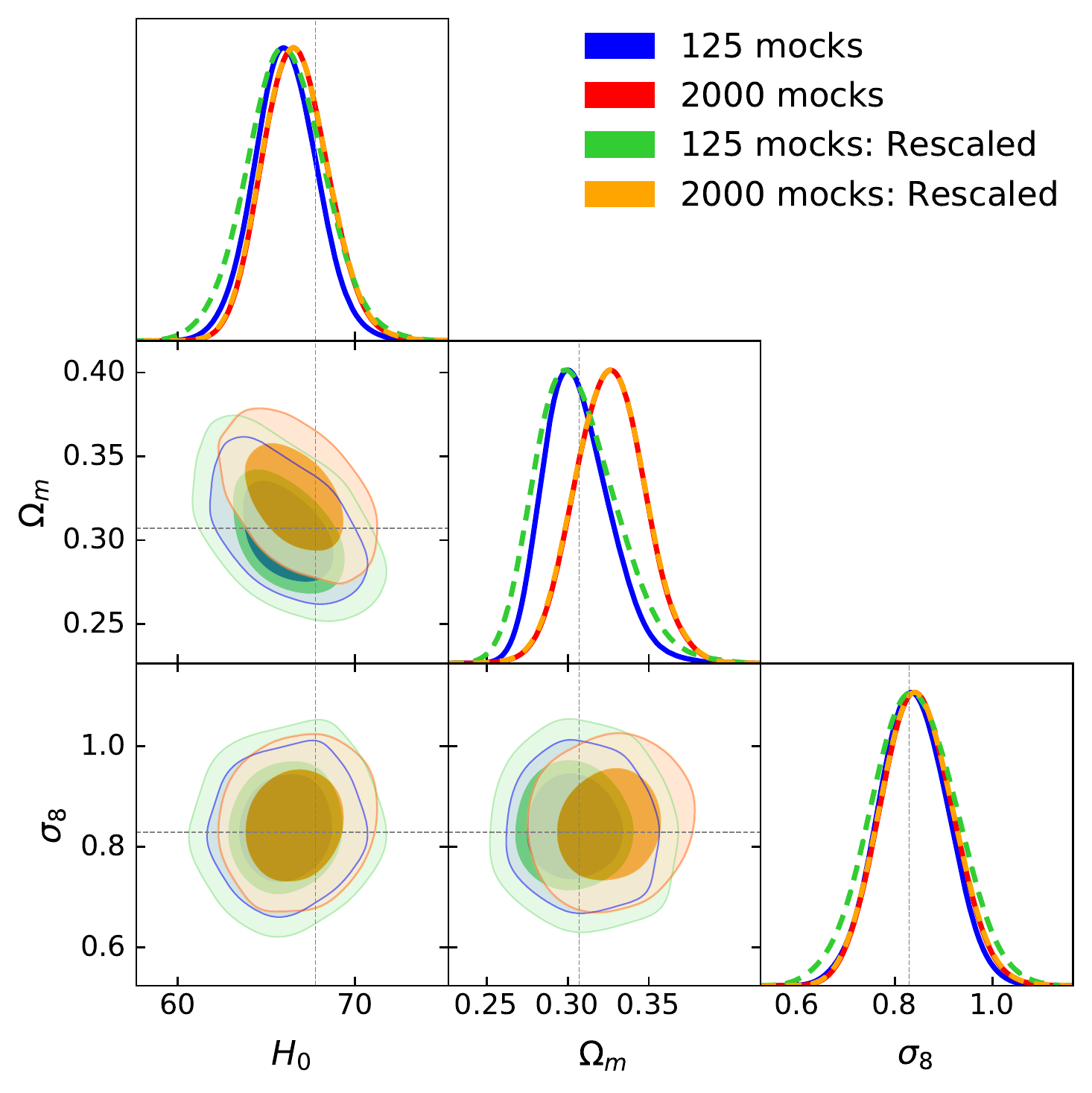} }}%
    \qquad
    \subfloat[12 Subspace Coefficients]{{\includegraphics[width=0.3\textwidth]{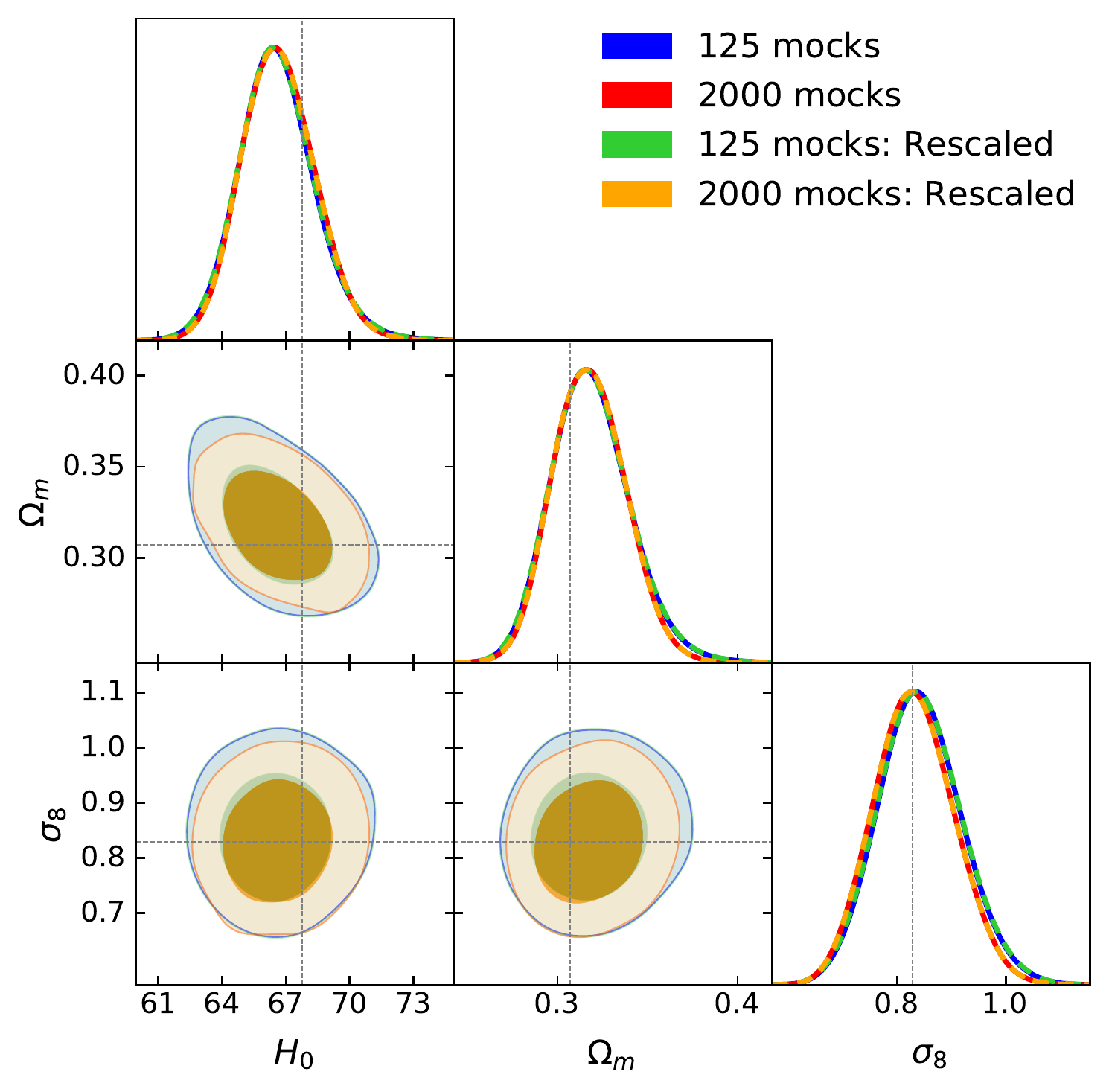} }}%
    \caption{Comparison of the posterior parameter contours from analyses using a sample covariance matrix created from 125 mocks (blue, full lines) or 2000 mocks (red, full lines), for various choices of data compression. Notably, there are significant shifts induced by precision matrix noise, but these are almost completely nulled by using $N_\mathrm{SV} = 12$, the value indicated by $\chi^2$ constraints. In green and yellow dashed lines we show the corresponding results including the $m_1$ rescaling factor of Eq.\,\ref{eq: param-cov-rescale} to account for precision matrix noise. With the rescaling, the posteriors with $N_\mathrm{mock} = 125$ and $N_\mathrm{mock} = 2000$ are not in tension, though the precision matrix noise leads to a significant loss of constraining power when many bins are used.}%
    \label{fig: precision-matrix-noise}%
\end{figure}

Perhaps the most important test is whether restricting to a small number of subspace coefficients allows us to use fewer mocks. To test this, we simply run the likelihood analysis with a single mock data-set and two sets of covariance matrices; one generated from 2000 mocks and one with eight times fewer (matching the factor by which we optimally compress the data). Note that we assume a Gaussian likelihood in both instances \resub{(apropos of the discussion in Sec.\,\ref{subsec: theory-noisy-precision})}. These results are shown in the first two data-sets of Fig.\,\ref{fig: precision-matrix-noise}, and we immediately note large shifts in the best-fit parameters for the uncompressed likelihood analysis, alongside an inflation of the parameter error bars, matching the predictions of Sec.\,\ref{subsec: theory-noisy-precision}. Notably, these shifts dramatically decrease as we compress the data, and, for $N_\mathrm{SV} = 12$ (the value suggested by the constraints on $\Delta\chi^2$ in Sec.\,\ref{subsec: template-bank}), these are insignificant. This is the main result of this paper; we can obtain robust estimates of cosmological parameters with just $\mathcal{O}(100)$ mocks when using optimal subspace projections.

Whilst the above discussion 
correctly shows the \resub{stochastic shifts} induced in the cosmological parameters from precision matrix noise, it does not accurately reflect real analyses, which usually account for this by inflating the parameter error bars by rescaling the Gaussian parameter covariances by the constant factor\footnote{It is important to note that we focus only on data-vectors that are not used to build the sample covariance. For this reason, we do not include the $M_2$ factor used in the mock catalog analyses of Refs.\,\citep{2014MNRAS.439.2531P,2017MNRAS.466.2242B}. Our $m_1$ factor is equivalent to $M_1^2$ in the notation of Ref.\,\citep{2017MNRAS.466.2242B}. \resub{Furthermore, we note the discussion at the end of Sec.\,\ref{subsec: theory-noisy-precision} regarding the applicability of the $m_1$ factor.}}
\beq\label{eq: param-cov-rescale}
    m_1 &=& \frac{1+B(N_\mathrm{bin}-N_\mathrm{param})}{1+A+B(N_\mathrm{param}+1)},
\eeq
defining
\beq
    A = \frac{2}{(N_\mathrm{mock}-N_\mathrm{bin}-1)(N_\mathrm{mock}-N_\mathrm{bin}-4)},\quad B = \frac{N_\mathrm{mock}-N_\mathrm{bin}-2}{(N_\mathrm{mock}-N_\mathrm{bin}-1)(N_\mathrm{mock}-N_\mathrm{bin}-4)}
\eeq
\citep{2014MNRAS.439.2531P}. This is more complex than that of Eq.\,\ref{eq: shift-covariance-factor} (which is equal to the numerator of $m_1$ \resub{minus one}), since noise in the covariance matrix inflates the observed parameter error bars in addition to giving a best-fit parameter shift. When rescaling the observed parameter covariances to account for best-fit variation, this must be accounted for, giving the reduced shift of Eq.\,\ref{eq: param-cov-rescale} \citep{2016MNRAS.456.4156K}. The $m_1$ numerator (equivalently, Eq.\,\ref{eq: shift-covariance-factor}) gives the necessary parameter covariance inflation relative to the $N_\mathrm{mock}\rightarrow\infty$ case, which is a good indicator of the efficacy of the subspace decomposition; for $N_\mathrm{mock} = 125$ and $N_\mathrm{param}$, we obtain $1+B(N_\mathrm{bin}-N_\mathrm{param}) \approx$ 4.3, 1.5, 1.0 for $N_\mathrm{bin} = 96,48,12$, whilst $m_1 \approx $ 3.0, 1.3, 1.0 for the same data. The results including $m_1$ are shown in the third and fourth data-sets in Fig.\,\ref{fig: precision-matrix-noise}; we note that the width of the one-dimensional histograms for the $N_\mathrm{mock}=125$, $N_\mathrm{bin}=96$ case is inflated by $\sim 2$x, 
as expected. Following the rescaling, the results for $N_\mathrm{mock} = 125$ and $2000$ are broadly consistent, though this is with a significant loss of precision, or equivalently, effective survey volume. \resub{As above, we note that strictly the likelihood should be replaced by a modified t-distribution to properly account for the effects of covariance matrix noise, thus the above rescaling, which is derived in the Gaussian limit, is not fully valid. Switching to the modified likelihood was found to have insignificant effect in our context, even for $N_\mathrm{mock} = 125$.}

\subsubsection{Changing the Fiducial Covariance}
\begin{figure}%
    \centering
    \subfloat[48 Subspace Coefficients]{{\includegraphics[width=0.45\textwidth]{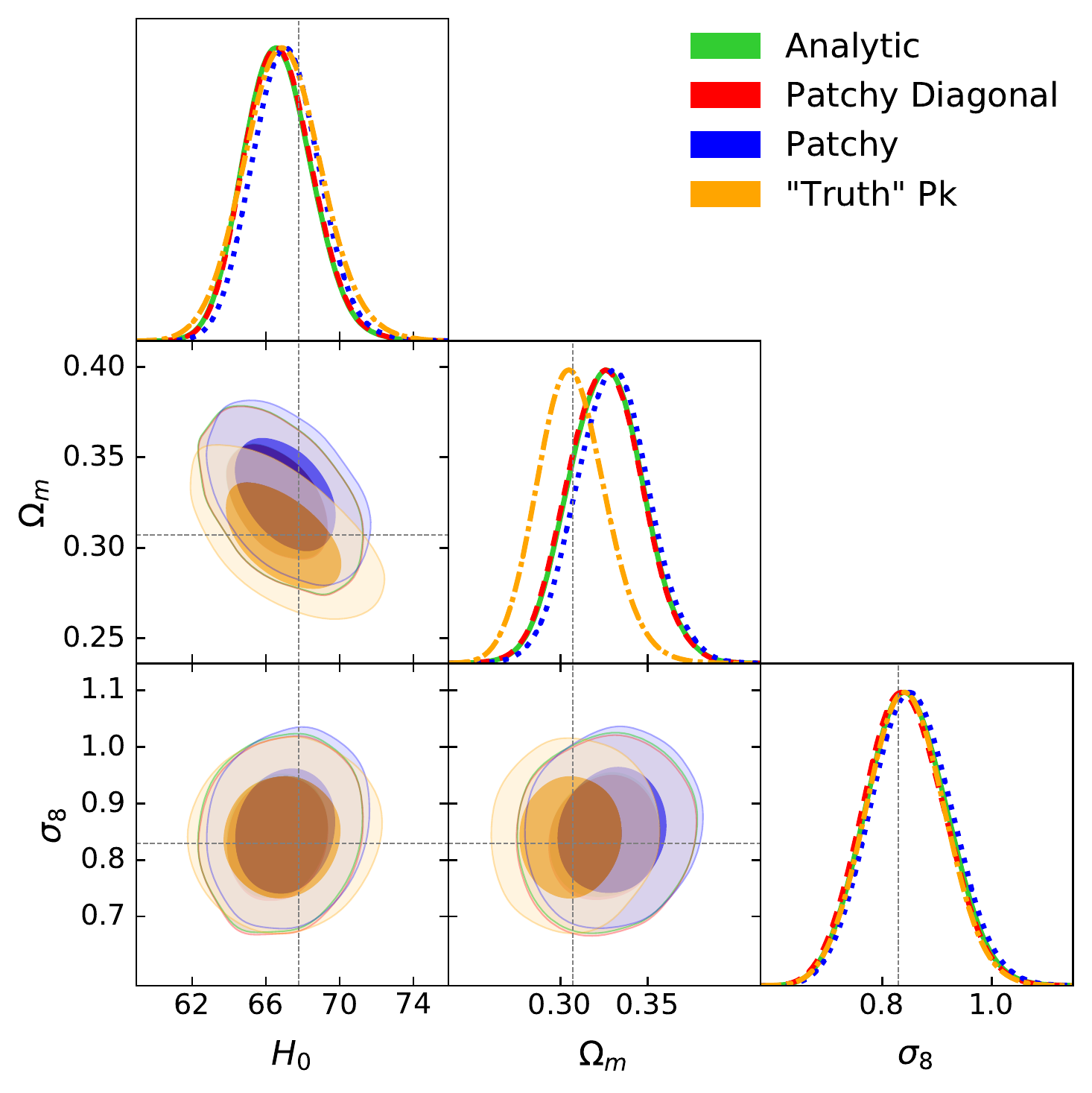} }}%
    \qquad
    \subfloat[12 Subspace Coefficients]{{\includegraphics[width=0.45\textwidth]{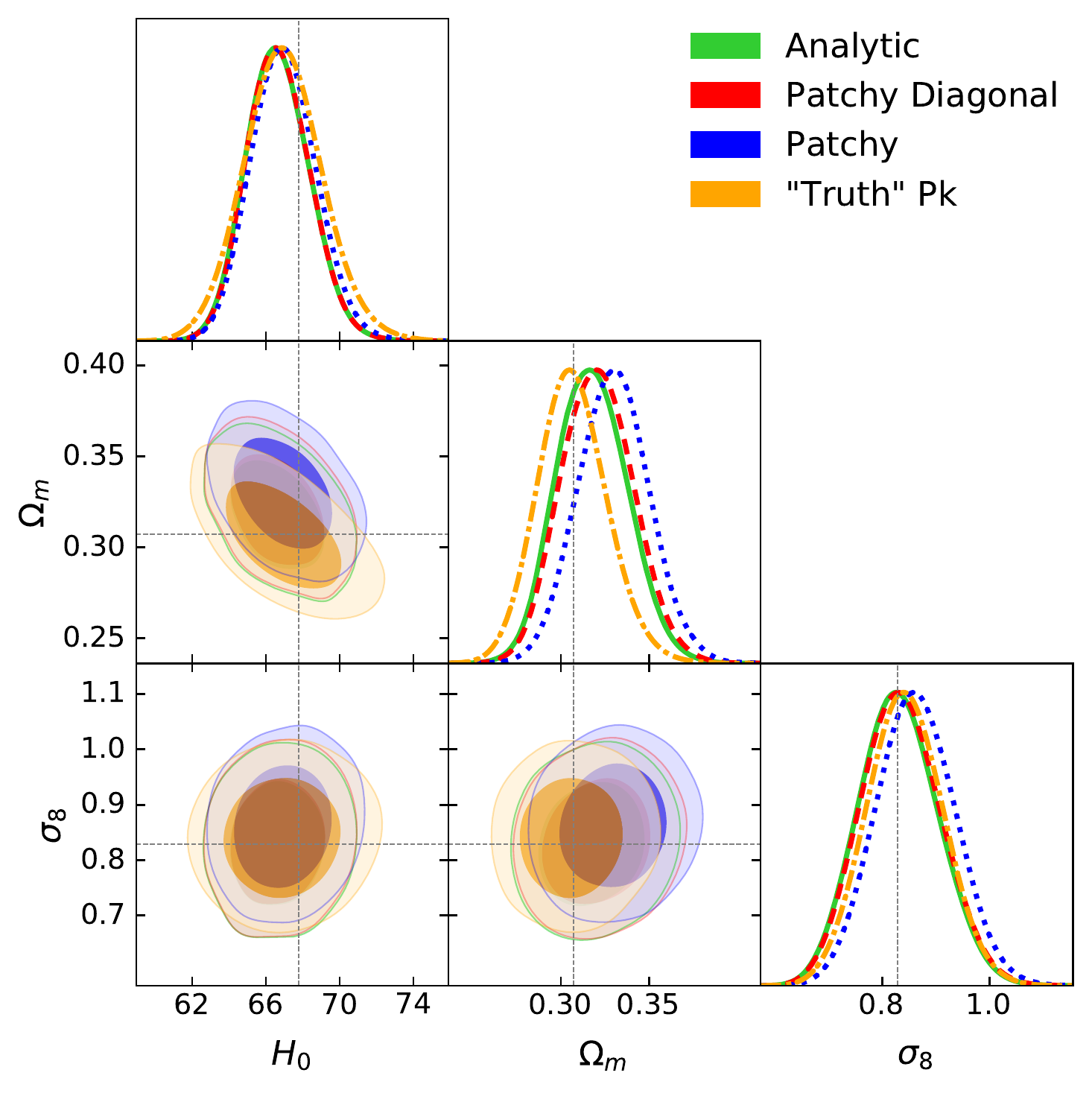} }}%
    \caption{Comparison of the parameter posteriors estimated using different choices of the fiducial covariance matrix, $\mathsf{C}$, to define the subspace decomposition. Results are plotted for the analytic covariance of Ref.\,\citep{2019arXiv191002914W} (green, full lines), the diagonal of the sample covariance from 2000 Patchy mocks (red, dashed lines), and the full Patchy covariance (blue, dotted lines). We additionally show the posterior from the mean-of-mocks analysis of Fig.\,\ref{fig: mean-of-mocks} (yellow, dot-dashed lines) and note that all likelihoods use the full 2000-mock Patchy covariance in the likelihood. For 48 SVs, there is little difference between choices of fiducial covariance, but some shifts for 12 SVs. We attribute this to noise in the SVD basis vectors induced by using noisy fiducial covariances (\textit{i.e.} Patchy).}%
    \label{fig: fiducial-cov}%
\end{figure}

An important hyperparameter is the fiducial covariance, $\mathsf{C}$, used to define the subspace. In Sec.\,\ref{subsec: theory-fiducial-choice}, it was stated that, whilst any (invertible) choice of fiducial covariance would lead to an unbiased estimate of cosmology when averaged over noise, noisy estimates, or those far from the true covariance, would lead to greater noise-induced shifts for small numbers of SVs. To test this, Fig.\,\ref{fig: fiducial-cov} compares the posterior predictions from a single mock power spectrum analyzed with three fiducial covariances: (1) the analytic covariance discussed above (with off-diagonal terms); (2) the diagonal of the Patchy covariance matrix from 2000 mocks; and (3) the full Patchy covariance matrix. In all cases, the data covariance is held constant.

Notably, we observe no significant differences between posteriors when using $N_\mathrm{SV} = 48$ but larger shifts with $N_\mathrm{SV} = 12$. In particular we note a \resub{significant shift} from using the Patchy fiducial covariance; this is attributed to the large off-diagonal noise present therein, which leads to less efficient subspace decompositions. This conclusion is strengthened by the results with the diagonal of the covariance matrix; whilst this does not accurately represent the true covariance (since window effects induce non-trivial mode-coupling and hence off-diagonal terms are present even if one assumes Gaussianity), it has low noise, and is consistent with the analytic prediction. We thus conclude that it is important to have a relatively smooth estimate of the fiducial covariance (though not necessarily one that matches the true covariance), else the low-SV results will be significantly affected by noise.

\subsubsection{Changing the Data Covariance}
\begin{figure}%
    \centering
    \subfloat[96-bin Power Spectrum]{{\includegraphics[width=0.3\textwidth]{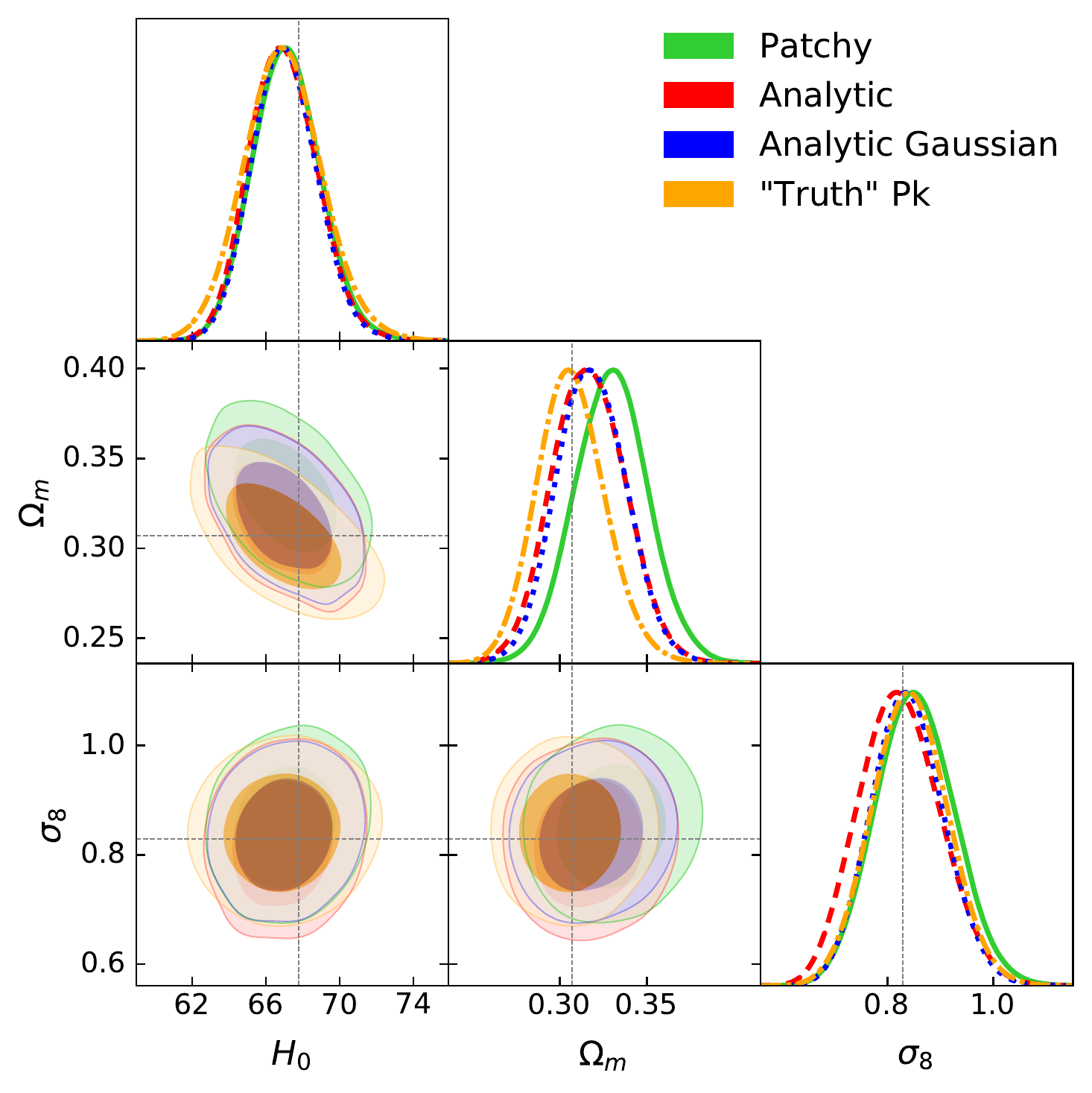} }}%
    \qquad
    \subfloat[48 Subspace Coefficients]{{\includegraphics[width=0.3\textwidth]{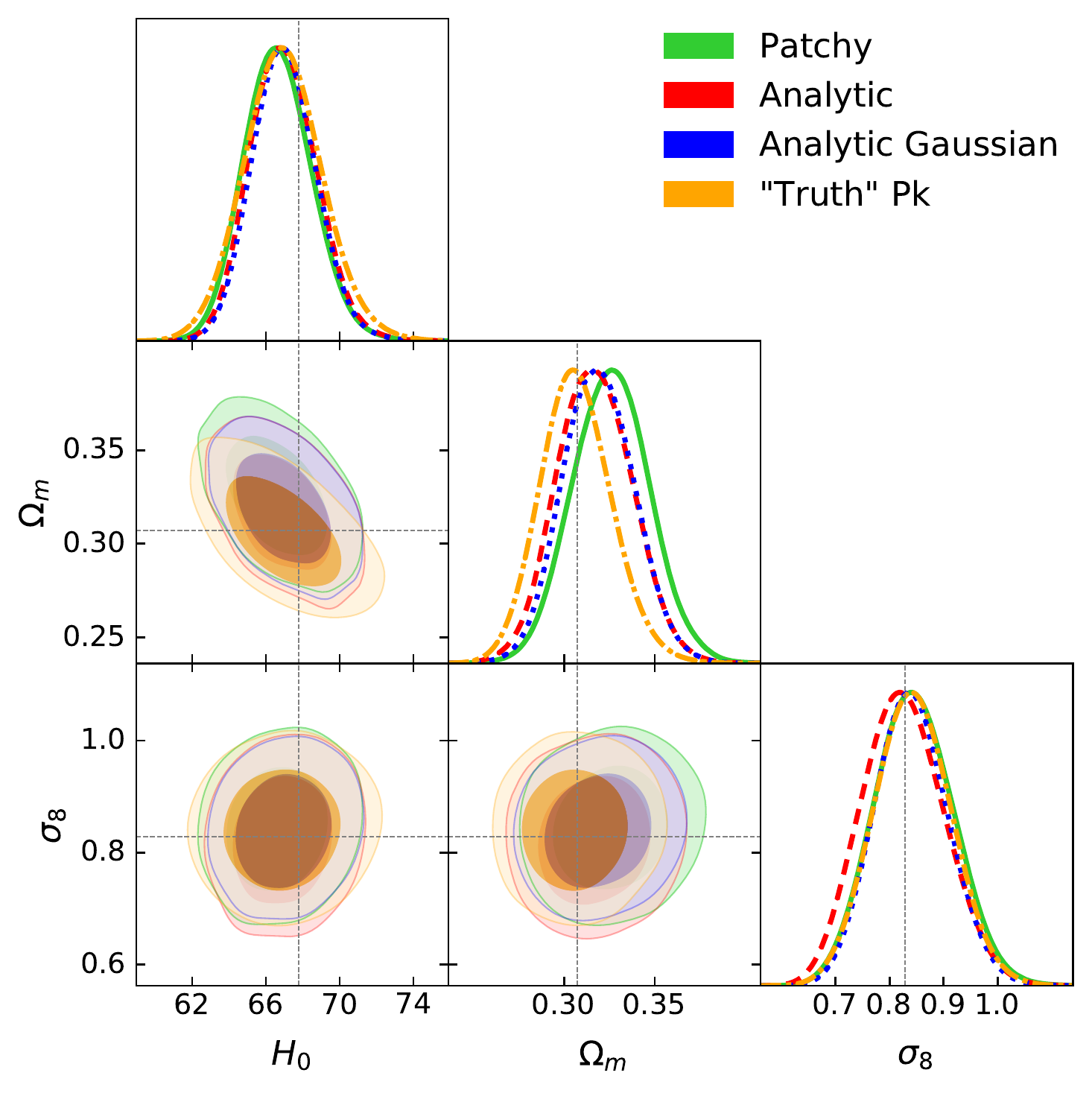} }}%
    \qquad
    \subfloat[12 Subspace Coefficients]{{\includegraphics[width=0.3\textwidth]{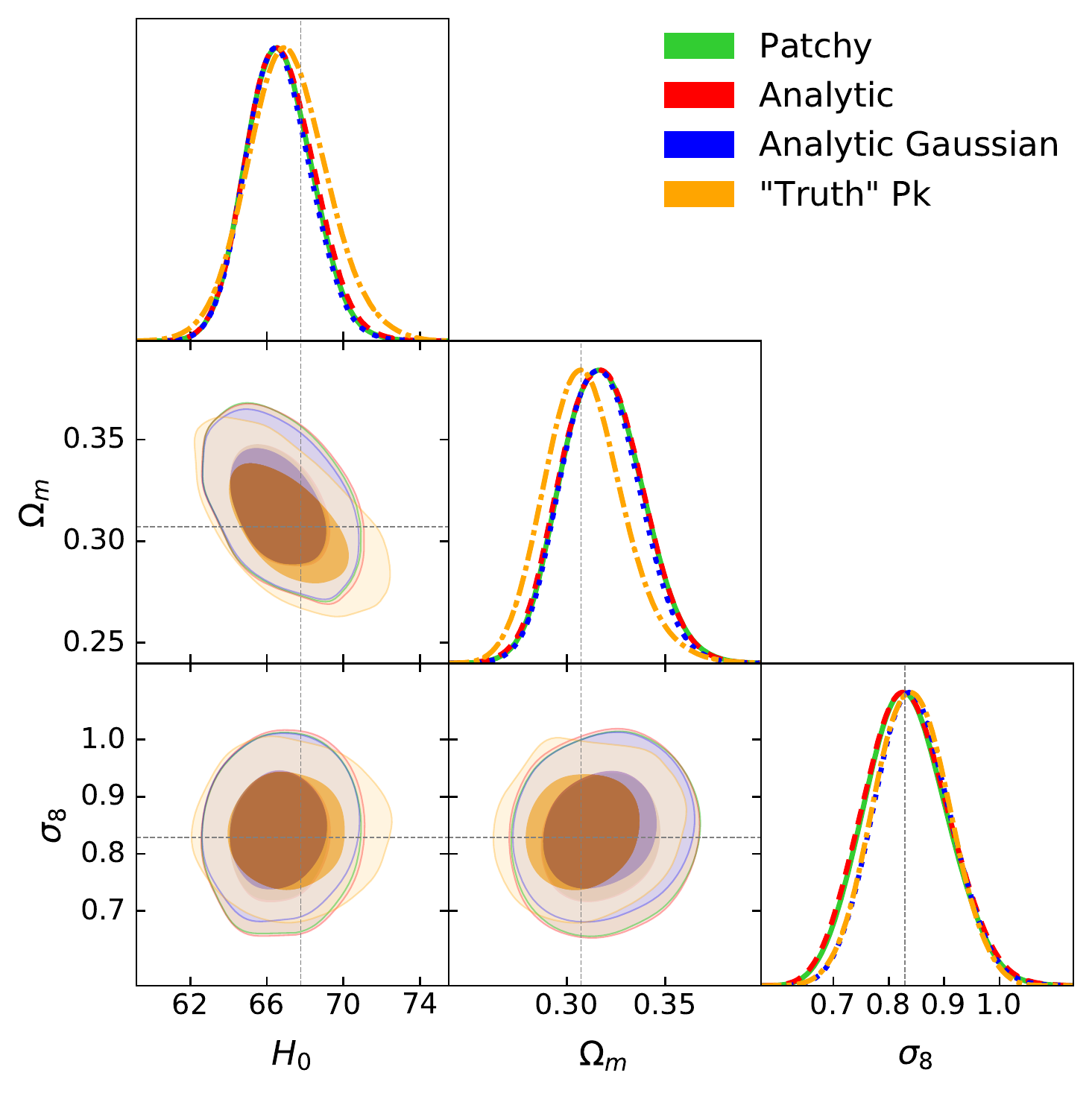} }}%
    \caption{Estimated posteriors from analyses using different choices of covariance matrix in the likelihood. Data is shown for the 2000-mock Patchy covariance (green, full lines), the analytic covariance of Ref.\,\citep{2019arXiv191002914W} (red, dashed lines), and an analytic Gaussian covariance (blue, dotted lines) alongside the mean covariance of Fig.\,\ref{fig: mean-of-mocks} (yellow, dot-dashed lines). In all cases, the fiducial covariance (used to define the subspace basis vectors) is set to the analytic covariance of Ref.\,\citep{2019arXiv191002914W}, and we analyze a single mock data set with differing compressions. Using 12 SVs, the posteriors are identical for each covariance, though there are larger noise-induced discrepancies when using greater numbers of bins.}%
    \label{fig: data-cov}%
\end{figure}

Can consistent results be obtained from analyses using different data covariances? This question extends beyond subspace analyses, but can be well-probed in our formalism, since we can separate out the effects of precision-matrix noise by reducing the number of basis coefficients. In Fig.\,\ref{fig: data-cov} we display the results from a selection of covariance matrices; the 2000-mock Patchy covariance, the analytic prescription of Ref.\,\citep{2019arXiv191002914W} (including trispectrum and super-sample terms) and an analytic Gaussian covariance. In the conventional analysis using 96 power spectrum bins, there are seen to be significant ($\sim 0.5\sigma$) shifts in $\Omega_m$ when different covariances are used, especially between Patchy and the analytic prescriptions.\footnote{These shifts are larger than those of Ref.\,\citep{Wadekar:2020rdu}; this is due to our restriction to a single data chunk, and the different random catalog used in the former work.} Notably, adding the trispectrum terms to the covariance does \textit{not} appear to induce a significant parameter shift. As $N_\mathrm{SV}$ decreases, these differences become small, and, by 12 SVs, we report no obvious deviations in the best-fit parameters when using different covariances.

From these results, one may conclude that, for a BOSS-like sample volume and tracer density, the trispectrum terms in the covariance do not alter the output parameter posteriors. Further, there is a bias induced by using the publicly-available Patchy covariance, which disappears as the number of bins is reduced. This can thus be attributed to residual noise in the covariance matrix, and must be carefully taken into account to avoid biasing the output cosmology, for example by using the subspace-based analysis.
These conclusions agree with the results of Ref.~\cite{Wadekar:2020rdu},
which analyzed the BOSS data using the perturbation theory covariance matrices \cite{2019arXiv191002914W}.

\section{Discussion}\label{sec: discussion}
\subsection{Dependence of Number of SVs on Priors and the Survey Volume}
In Sec.\,\ref{sec: template-creation}, we found that, for the BOSS sample used in this work, 12 SVs were needed to ensure that the mean subspace $\chi^2$ was consistent with the usual power spectrum $\chi^2$ to within $\chi^2_\mathrm{min} = 0.1$. This is not a general statement, since it depends on the fiducial covariance matrix, the cosmological model and our choice of priors. In the above, we opted to use broad priors implying a lack of knowledge of the posterior; tighter priors lead to smaller variation of $\chi^2$ across the parameter manifold, thus requiring fewer SVs. A simple test of this is to repeat the SVD for power spectrum samples drawn from the \textit{posterior} rather than the prior template bank. This is simply done, and represents the minimum number of SVs which can be safely used, since it is the number one would obtain if they started with full knowledge of the posterior space. In this case, we find that only 8 subspace coefficients are required, rather than 12. This is \textit{less} than the number of model parameters, indicating that the parameters are partially degenerate. \resub{A hybrid approach might also be possible; \textit{i.e.} one could approximate the posterior as a multivariate Gaussian and draw template banks instead from this distribution. Whilst this would result in basis vectors that are more tailored to the posterior peak and thus somewhat fewer SVs, it will fail if the posterior is not well-modeled as a Gaussian (noting that a Gaussian posterior is not necessarily implied by a Gaussian likelihood). For the the sake of generality, we have not implemented such an approach here.}

The required number of SVs also has dependence on the observational volume. If the survey size is increased by a factor $f$, we expect the covariance to fall by a factor $f$, and thus the SVD matrix $D_\alpha$ to scale as $f^{1/2}$. From Eq.\,\ref{eq: N-sv-lim}, we may mimic this by choosing $N_\mathrm{SV}$ such that the cumulative signal-to-noise of excluded SVs is less than $\chi^2_\mathrm{min}/f$ rather than $\chi^2_\mathrm{min}$. For the priors considered herein, increasing the survey volume by a factor of 10 (making it comparable to the DESI volume), whilst making the unrealistic assumption that there is still only one tomographic bin, the number of required SVs increases from 12 to 16. This increase is due to the non-linear parameter dependencies becoming more important (with respect to noise), though we note it to still be a small number compared to the 96 power spectrum bins, due to the steepness of the $D_\alpha$.


\subsection{Application to Bispectra}
Whilst the main focus of this paper has been the galaxy power spectrum, the subspace decomposition is fully general and can be applied to any observable, given a theory model and a set of priors. Of particular interest is the galaxy bispectrum, since this statistic generically has a large number of bins, which has limited its applicability in previous mock-based approaches.\footnote{As an example, Ref.\,\citep{2017MNRAS.465.1757G} had to use broad $k$-bins to ensure that the covariance could still be estimated from the 2048 available mock catalogs.} Any formalism that is able to substantially compress the bispectrum, whilst retaining its information content, is thus of great importance, allowing mock-based analyses to take place in reasonable computation times.\footnote{See Ref.\,\citep{2000ApJ...544..597S} for an approximate bispectrum compression method using bispectrum eigenmodes.} To test the applicability of our method to the bispectrum, we may simply ask the question: how many SVs are needed to reproduce the bispectrum likelihood to within $\Delta\chi^2 = 0.1$?

For this forecast, we consider a simplified scenario; a survey with the effective volume and redshift of the largest BOSS data-chunk, but with a periodic box geometry. We assume the same power spectrum model as above (one-loop effective field theory), keeping the $k$-binning constant (\textit{i.e.} with 48 $k$ bins for each of the monopole and quadrupole). For the (redshift-space monopole) bispectrum, we use tree-level theory as in Ref.\,\citep{2019JCAP...11..034C}, for $k$ in $[0.01,0.15]\hMpc$. This gives a total of $2135$ bispectrum bins and $96$ power spectrum bins. To generate the subspace basis vectors we require also a fiducial model for the joint covariance (Sec.\,\ref{subsec: template-bank}); for this we assume a Gaussian covariance for both observables (matching that of Ref.\,\citep{2019JCAP...11..034C}), noting that the exact choice of fiducial covariance is of limited importance in the analysis. For simplicity, we assume the power spectrum and bispectrum to be uncorrelated, \textit{i.e.} that they are observed in separate regions of the sky. By instead inserting a Gaussian model for the cross-covariance \cite{Smith:2007sb}, we have shown that this assumption does not have a significant impact on the observed $\chi^2$, or number of SVs.

\begin{figure}
    \centering
    \includegraphics[width=0.6\textwidth]{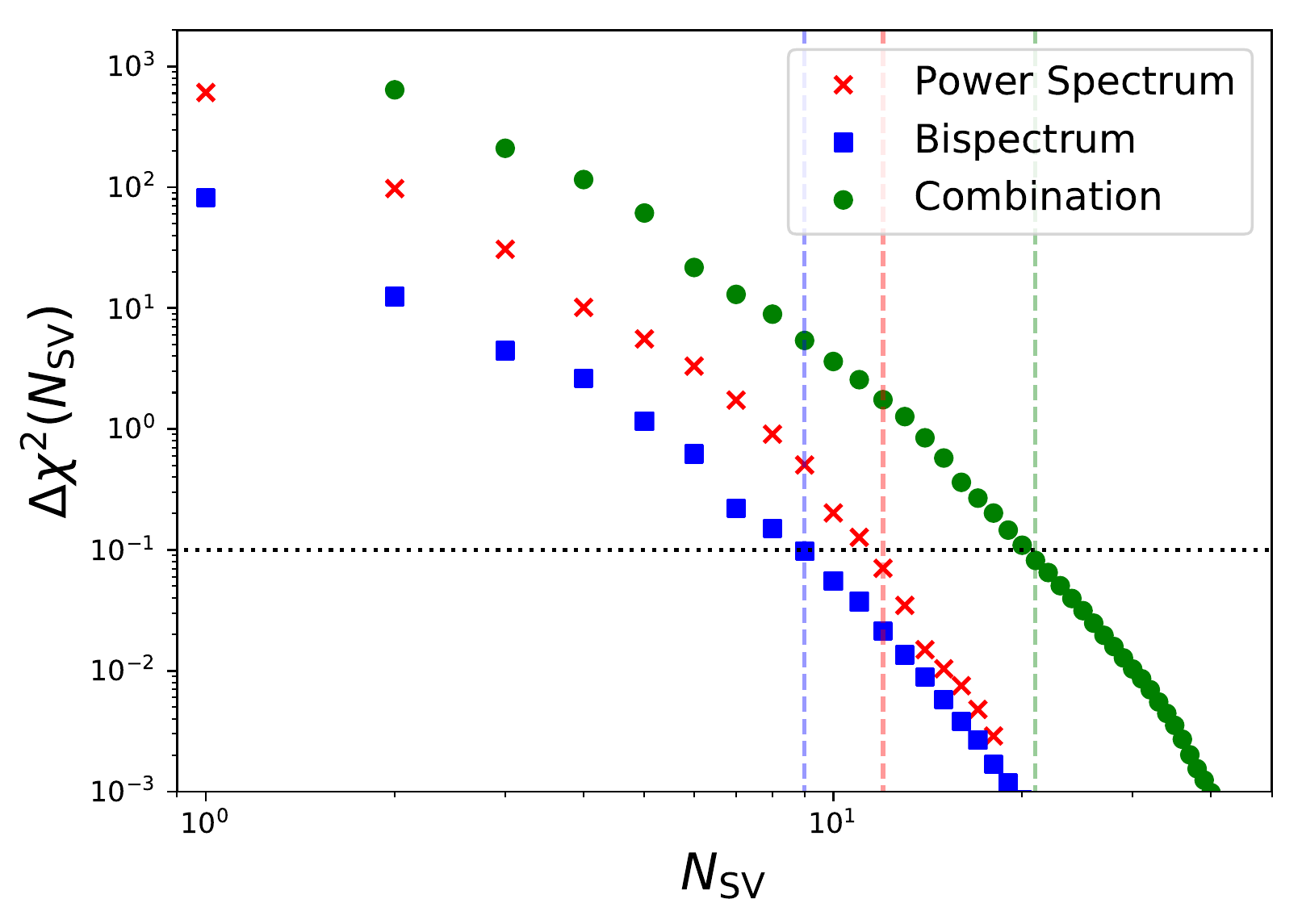}
    \caption{Impact on $\chi^2$ from using a finite number of basis vectors in the likelihood for mock analyses using the power spectrum, the bispectrum, and the combination of two probes (as described in the main text). For each case we plot the difference between the true $\chi^2$ (from all $N_\mathrm{bin}$ bins) and that using $N_\mathrm{SV}$ SVs. The horizontal line indicates $\Delta\chi^2 = 0.1$, \textit{i.e.} beyond this point, the sum of all remaining SVs contribute less than $0.1$ to the total log-likelihood. For the power spectrum analysis, 12 basis vectors are required to reach this limit (compared to 96 bins), whilst for the bispectrum, only 9 are needed (compared to 2135 bins). In combination, one needs 21 SVs, though this is an upper bound as the two observables are assumed to be uncorrelated.}
    \label{fig: pkbk-svs}
\end{figure}

A set of $10^4$ template bank samples is then computed from the prior, and the SVD performed. As in Sec.\,\ref{subsec: template-bank}, the output SVs can be used to assess the impact on the prior-averaged $\chi^2$ from including only a subset of all bins; in Fig.\,\ref{fig: pkbk-svs}, we plot the $\chi^2$ deficit for analyses using (a) only the power spectrum, (b) only the bispectrum, and (c) their combination. As previously noted, there is a steep dependence of $\Delta\chi^2$ on the number of SVs, with negligible contributions from the higher basis vectors. As before, we consider the optimal $N_\mathrm{SV}$ to be the minimum number which yields a total (prior-averaged) $\chi^2$ error below $0.1$; this is found to be 12 for the power spectrum, 9 for the bispectrum, and 21 for their combination.\footnote{\new{If we restrict only to modes with $k<0.1\hMpc$ for the tree-level bispectrum, we find that only 5 (17) SVs are needed for the bispectrum-only (combined) analysis.}} Further, any correlations between observables are expected to reduce the combined number. The efficacy of this method is impressive; we can compress a combined data-vector with $2231$ bins into just $21$ (even in the presence of non-linearities in the likelihood), which would allow easy analysis with $\mathcal{O}(100)$ mocks. \resub{This has further implications for the \textit{measurement} of bispectra from data; by including the optimal basis vectors directly in the bispectrum estimators, the relevant compressed statistics can be computed in far fewer operations than before; an important quality given the non-trivial computation time required for bispectrum estimation.}

\new{Similar analysis is possible for the combination of full-shape analyses, such as that presented herein, with BAO information from reconstructed power spectra (as in Ref.\,\citep{2020JCAP...05..032P}). This is straightforward to test; we simply generate a template bank of the power spectrum multipoles coupled with the Alcock-Paczynski parameters, $\vec\alpha=\{\alpha_\parallel, \alpha_\perp\}$, and apply the SVD, supplementing the fiducial covariance with the measured joint-covariance of $\vec\alpha$ and multipoles discussed in Ref.\,\citep{2020JCAP...05..032P}. Here, we find that only a single additional extra SV is required to accurately approximate the (significantly more constraining) $\chi^2$.}



\section{Summary}\label{sec: summary}
In this work, we have introduced a formalism to reduce the dimensionality of cosmological observables by linearly projecting the observables into subspaces \resub{that maximize the prior-averaged $\chi^2$}. Given only a theoretical model and a set of priors, a set of model predictions can be computed and used to define a quasi-Euclidean metric space via a singular value decomposition. This provides a natural compression for the observables; by restricting to the first $N_\mathrm{SV}$ basis vectors of the subspace, the dimensionality of the data is significantly reduced whilst the likelihood remains almost unchanged. \resub{Whilst we assume the likelihood itself to be Gaussian, it can be non-Gaussian in the parameters themselves (and indeed multi-modal), since our decomposition is blind to the model parametrization.} Observables such as the power spectrum and bispectrum can be represented with $\mathcal{O}(10)$ subspace coefficients, incurring $\chi^2$ errors of $< 0.1$ over the broad prior manifold, even in the presence of non-linearities in the likelihood. Whilst the method has some dependence on an assumed fiducial covariance, this does not bias the inference, and requires only a simple (low-noise) estimate.

The principal appeal of this concerns covariance matrices; likelihood analyses are inherently sensitive to precision matrix noise induced by the use of finite numbers of mocks, which is drastically reduced by data compression. Indeed, we prove that restricting to low-dimensional subspaces significantly reduces the induced parameter \resub{shifts}, whilst keeping the noise-averaged constraints unchanged (except for a slight increase in parameter covariances, which is shown to be negligible in practice). Indeed, due to the \resub{posterior inflation required} to ensure that the estimators remain appropriate in the presence of noise, 
ur subspace projection gives more precise parameter constraints when the number of mocks is small.

Our method is validated by application to mock BOSS DR12 power spectrum multipoles. In particular, we show that the 96-bin power spectrum data can be robustly compressed to a set of 12 subspace coefficients, allowing for accurate parameter estimations using only $125$ mocks. The required number of coefficients has only weak dependence on the prior space; adding maximally restrictive priors would allow the number of coefficients to be reduced to only 8, in our 10-parameter model. \new{Whilst we have assumed a simple $\Lambda$CDM model in this work, this is by no means a restriction; the formalism can apply to any extension for which model spectra can be evaluated and further include arbitrary parameters describing systematics. Whilst one should recompute the basis vectors given a new cosmological model, this is not a major concern, since creating the template bank requires computing far fewer spectra than needed for a well-converged MCMC analysis.}


It is useful to compare our methodology to the SVD-based techniques used to de-noise the power spectrum and bispectrum covariance matrices in Refs.\,\cite{Eisenstein:1999jg,Scoccimarro:2000sn,Gaztanaga:2005ad}.
In this approach one diagonalizes the covariance matrix and uses only the highest signal-to-noise ($S/N$) eigenvalues in the likelihood analysis. In contrast to $S/N$, we use $\Delta \chi^2$ from variations of all
relevant parameters as a criterion to define the subspace projections, which allows us to better capture the eigenmodes that are sensitive to features in the power spectrum shape. Indeed, we have seen that our method can fully extract the information encoded in the BAO wiggles, which contribute very little to the overall $S/N$, but dominate the distance constraints. 

With the addition of just a small number of extra coefficients, the method can also apply to larger volumes (where non-linearities in the likelihood become more important, and thus Fisher-based compressions \resub{become suboptimal}), as well as more complex statistics, such as bispectra or combination with BAO constraints. For the BOSS sample analyzed herein, we need only 12 SVs to encapsulate $96$ power spectrum bins; this rises to $16$ for a ten-times larger survey (like DESI), and only a further eight are required to incorporate the $\sim 2000$ bin bispectrum. An additional possibility is for projected statistics such as lensing; the large number of bins across different redshifts in 3x2pt analyses could be substantially reduced using SVs. In general, the exact number of subspace coefficients depends both on the chosen parameter and experimental set-up (e.g., survey redshift and volume) and should thus be computed separately for each experiment, but this is a trivial operation to implement once the SVD has been performed.

The methods developed herein provide a useful sandbox in which to investigate an important question in cosmology: how parameter estimates depend on power spectrum covariances. Focusing on the largest data-chunk of BOSS, we find that the inclusion of off-diagonal trispectrum terms in the covariance does not affect the parameter estimates, though this will generically depend on the shot-noise and $k$ binning. Further, in the uncompressed likelihood, there is a $\sim 0.5\sigma$ stochastic shift in the best-fit posteriors for $\Omega_m$ using a covariance drawn from MultiDark-Patchy mocks compared to that using analytic covariances, away from the true cosmology and in the direction of increasing tension with \textit{Planck}. In contrast, using a small number of subspace coefficients, all covariances yield consistent results, indicating that (a) there is residual noise in the mock-based covariance that shifts the inferred cosmological parameters even if 2000 mock catalogs are used, and (b) analytic prescriptions can be accurately applied. Subspace projection provides a simple way of ameliorating this, through limiting the effects of covariance matrix noise. \resub{Furthermore, the ability to perform analyses using significantly fewer mocks paves the way to allowing the covariance matrix to be parameter-dependent; the computational cost of performing repeated analyses gradually updating the covariance matrix becomes far more manageable.}

Approaches such as that of this work are vital when using statistics beyond the power spectrum. In particular, bispectrum analyses should no longer be considered mock-limited, opening up new and exciting avenues into the exploration of cosmological data-sets. The future grows brighter for higher-order statistics.

\begin{acknowledgments}
We thank Jay Wadekar and Roman Scoccimarro for providing us with analytic covariance matrices and sharing a draft of Ref.\,\citep{Wadekar:2020rdu}. 
We additionally thank \resub{Alex Hall, Alan Heavens, Dragan Huterer}, David Spergel, \resub{Masahiro Takada}, Jay Wadekar, \resub{Ben Wandelt}, Martin White, \resub{and the anonymous referee} for comments on the manuscript. OHEP acknowledges funding from the WFIRST program through NNG26PJ30C and NNN12AA01C and thanks the Max Planck Institute for Astrophysics for hospitality when this work was being finalized. MI is partially supported by the Simons Foundation’s \textit{Origins of the Universe} program. MS acknowledges support from the Corning Glass Works Fellowship and the National Science Foundation. MZ is supported by NSF grants PHY-1820775 the Canadian Institute for Advanced Research (CIFAR) Program on Gravity and the Extreme Universe and the Simons Foundation Modern Inflationary Cosmology initiative.

Funding for SDSS-III has been provided by the Alfred P. Sloan Foundation, the Participating Institutions, the National Science Foundation, and the U.S. Department of Energy Office of Science. The SDSS-III web site is http://www.sdss3.org/.

SDSS-III is managed by the Astrophysical Research Consortium for the Participating Institutions of the SDSS-III Collaboration including the University of Arizona, the Brazilian Participation Group, Brookhaven National Laboratory, Carnegie Mellon University, University of Florida, the French Participation Group, the German Participation Group, Harvard University, the Instituto de Astrofisica de Canarias, the Michigan State/Notre Dame/JINA Participation Group, Johns Hopkins University, Lawrence Berkeley National Laboratory, Max Planck Institute for Astrophysics, Max Planck Institute for Extraterrestrial Physics, New Mexico State University, New York University, Ohio State University, Pennsylvania State University, University of Portsmouth, Princeton University, the Spanish Participation Group, University of Tokyo, University of Utah, Vanderbilt University, University of Virginia, University of Washington, and Yale University.
\end{acknowledgments}

\appendix

\section{Increase in Parameter Covariance from Subspace Projection}\label{appen: fisher-matrices}
We here demonstrate that the shift in the parameter covariance (Eq.\,\ref{eq: shift-parameter-cov}) is positive semidefinite, implying that the restriction to a subspace increases the covariance on average. For convenience, and without loss of generality, we adopt the parameter set $\vec\psi$ in which the Fisher matrix is diagonal (this can always be found by diagonalization). We further rotate the coefficients $c(\vec\psi)\rightarrow \tilde{c}(\vec\psi) = \mathcal{C}_D^{-1/2}c(\vec\psi)$ such that the subspace metric is diagonal. In this basis, the full Fisher matrix is simply;
\beq\label{eq: diagonal-fisher}
    \mathcal{F}_{ij} = \frac{\delta^K_{ij}}{\sigma_i\sigma_j} \quad \Rightarrow \quad \Phi_{ij} = \delta^K_{ij}\sigma_i\sigma_j, \quad \Delta\Phi_{ij} = - \sigma^2_i\sigma^2_j\Delta\mathcal{F}_{ij},
\eeq
where $\sigma^2_i$ is the variance of parameter $\psi_i$. The Fisher matrix in the subspace defined by $N_\mathrm{SV}\leq N_\mathrm{bin}$ SVs is given by
\beq
    \hat{\mathcal{F}}_{ij} = \sum_{\alpha=1}^{N_\mathrm{SV}}\sum_{\beta=1}^{N_\mathrm{SV}}\frac{\partial \tilde{c}_\alpha(\vec\psi^*)}{\partial \psi_i}\frac{\partial \tilde{c}_\beta(\vec\psi^*)}{\partial \psi_j} \leq \sqrt{\sum_{\alpha=1}^{N_\mathrm{SV}}\left(\frac{\partial \tilde{c}_\alpha(\vec\psi^*)}{\partial \psi_i}\right)^2\sum_{\beta=1}^{N_\mathrm{SV}}\left(\frac{\partial \tilde{c}_\beta(\vec\psi^*)}{\partial \psi_j}\right)^2}
\eeq
using the Cauchy-Schwarz inequality in the second line. Since $N_\mathrm{SV}\leq N_\mathrm{bin}$ and the derivatives are real, each component of the sum is positive, hence
\beq
    \hat{\mathcal{F}}_{ij} \leq \sqrt{\sum_{\alpha=1}^{N_\mathrm{bin}}\left(\frac{\partial \tilde{c}_\alpha(\vec\psi^*)}{\partial \psi_i}\right)^2\sum_{\beta=1}^{N_\mathrm{bin}}\left(\frac{\partial \tilde{c}_\beta(\vec\psi^*)}{\partial \psi_j}\right)^2} = \frac{1}{\sigma_i\sigma_j},
\eeq
and thus
\beq
    \Delta\mathcal{F}_{ij} \equiv \hat{\mathcal{F}}_{ij} - \mathcal{F}_{ij} \leq \frac{1}{\sigma_i\sigma_j}\left(1 - \delta_{ij}^K\right).
\eeq
Contracting this with an arbitrary vector $x$, one can easily show this to be negative semidefinite;
\beq
    \sum_{ij}x_i\Delta\mathcal{F}_{ij} x_j = \left[\sum_i \left(\frac{x_i}{\sigma_i}\right)^2 - \sum_i \frac{x_i^2}{\sigma_i^2}\right] \leq 0,
\eeq
where we have again invoked the Cauchy-Schwarz inequality. From Eq.\,\ref{eq: diagonal-fisher}, the parameter covariance shift $\Delta\Phi$ is thus positive semidefinite.

\section{Parameter Shifts from Noisy Data and Noisy Covariances}\label{appen: param-shifts}
We here derive the expected \resub{stochastic shift} in the best fit parameters from both noisy data and a noisy covariance matrix estimate. Much of these derivations parallel Refs.\,\citep{2013PhRvD..88f3537D,2013MNRAS.432.1928T}, but we include them here for completeness. To begin, consider the observed $\chi^2$ using noisy precision matrix $\hat{\Psi} = \Psi+\delta\Psi$ ($\Psi \equiv \mathcal{C}_D^{-1}$ in the notation of Sec.\,\ref{sec: methodology}) and data $\hat{c} = c(\vec\theta^*) + \hat{n}$ where $\vec\theta^*$ are the true parameters and $n$ is some vector of noise (with each element drawn from a unit Gaussian for a diagonal subspace metric);
\beq
    \hat{\chi}^2(\vec\theta) = \left(c_\alpha(\vec\theta)-\hat{c}_\alpha\right)\hat{\Psi}_{\alpha\beta}\left(c_\beta(\vec\theta)-\hat{c}_\beta\right),
\eeq
assuming implicit index summation for brevity. The best-fit parameter vector, $\hat{\vec\theta}$, is found by minimization;
\beq\label{eq: noisy-chi2-min}
    \left.\frac{\partial \hat{\chi}^2}{\partial\theta_i}\right|_{\hat{\vec\theta}} = 0 \quad \Rightarrow \quad \left.\frac{\partial c_\alpha}{\partial\theta_i}\right|_{\hat{\vec\theta}}\hat{\Psi}_{\alpha\beta}\left(c_\beta(\hat{\vec\theta})-\hat{c}_\beta\right) = 0
\eeq
for all $i$. Assuming the deviation from the true parameters, $\vec\theta^*$, to be small, we can Taylor expand to first order in $\delta\vec\theta\equiv\hat{\vec\theta}-\vec\theta^*$;
\beq
    c_\alpha(\hat{\vec\theta}) = c_\alpha(\vec\theta^*) + \left.\frac{\partial c_\alpha}{\partial\theta_i}\right|_{\vec\theta^*}\delta\theta_i\,, \quad \left.\frac{\partial c_\alpha}{\partial\theta_i}\right|_{\hat{\vec\theta}} =  \left.\frac{\partial c_\alpha}{\partial\theta_i}\right|_{\vec\theta^*} +  \left.\frac{\partial^2 c_\alpha}{\partial\theta_i\partial\theta_j}\right|_{\vec\theta^*}\delta\theta_j.
\eeq
Inserting these into Eq.\,\ref{eq: noisy-chi2-min} and simplifying, we obtain
\beq\label{eq: parameter-error-tmp}
    \left\{\frac{\partial c_\alpha}{\partial\theta_i}\left(\Psi_{\alpha\beta}+\delta\Psi_{\alpha\beta}\right)\frac{\partial c_\beta}{\partial\theta_j} -\frac{\partial^2 c_\alpha}{\partial\theta_i\partial\theta_j}\Psi_{\alpha\beta}\hat{n}_\beta \right\}\delta\theta_j = \frac{\partial c_\alpha}{\partial\theta_i}\left(\Psi_{\alpha\beta}+\delta\Psi_{\alpha\beta}\right)\hat{n}_\beta,
\eeq
implicitly assuming all terms to be evaluated at $\vec\theta^*$. A further simplification is obtained by noting that the term in curly parentheses in Eq.\,\ref{eq: parameter-error-tmp}, is simply the Fisher matrix of a noisy realization, $\hat{\mathcal{F}}_{ij} = \mathcal{F}_{ij}+\delta\mathcal{F}_{ij}$. The parameter shift is thus
\beq\label{eq: parameter-error}
    \delta\theta_i = \left(\mathcal{F}+\delta\mathcal{F}\right)^{-1}_{ij}\frac{\partial c_\alpha}{\partial\theta_j}\left(\Psi_{\alpha\beta}+\delta\Psi_{\alpha\beta}\right)\hat{n}_\beta,
\eeq
agreeing with standard results (e.g., Eq.\,23 of Ref.\,\citep{2013PhRvD..88f3537D}). To simplify this further, note that
\beq\label{eq: fisher-variation}
    \left(\mathcal{F}+\delta\mathcal{F}\right)^{-1}_{ij} = \Phi_{ij}-\Phi_{ik}\delta\mathcal{F}_{kl}\Phi_{lj}
\eeq
at first order, where the $\Phi = \mathcal{F}^{-1}$ is the standard parameter covariance.

A number of well-known results are apparent from Eq.\,\ref{eq: parameter-error}: (a) the best-fit parameters are shifted by noisy data; (b) the amplitude of the shift is proportional to the parameter covariance; (c) noise on the precision matrix gives an additional shift in parameters, though is only present when the data is itself noisy. Note we have made no assumptions thus far on the number of SVs used in the $\alpha$, $\beta$ summation, thus the above results indicate that, for noise-free data, we will \textit{never} have a bias from using too few SVs. 

To quantify the extent of these parameter shifts, it is easiest to consider two regimes separately, first setting $\delta\Psi = 0$. The covariance of $\delta\vec\theta$ is then
\beq
    \av{\left.\delta\theta_i\right|_{\delta\Psi=0}\left.\delta\theta_j\right|_{\delta\Psi=0}} = \Phi_{ik}\Phi_{jl}\frac{\partial c_\alpha}{\partial\theta_k}\frac{\partial c_\beta}{\partial\theta_l}\av{\hat{n}_\alpha\hat{n}_\beta} \equiv \Phi_{ik}\Phi_{jl}\mathcal{F}_{kl} = \Phi_{ij},
\eeq
using $\av{\hat{n}_\alpha\hat{n}_\beta} = \mathcal{C}_{D,\alpha\beta} = \Psi_{\alpha\beta}^{-1}$ and ignoring noise contributions to $\delta\mathcal{F}_{ij}$. This is a standard result; the covariance in the parameter shift $\delta\vec\theta$ is just the inverted Fisher matrix. Combining Eqs.\,\ref{eq: parameter-error}\,\&\,\ref{eq: fisher-variation}, we find that the precision matrix noise induces an \textit{extra} shift in the best-fit parameter vector
\beq\label{eq: cov-noise-shift}
    \Delta\theta_i \equiv \delta\theta_i - \left.\delta\theta_i\right|_{\delta\Psi = 0} = 
    \left[\Phi_{ij}\hat{n}_\beta\frac{\partial c_\alpha}{\partial\theta_j} - \Phi_{ik}\Phi_{jl}\Psi_{\gamma\delta}\hat{n}_\gamma\frac{\partial c_\alpha}{\partial\theta_k}\frac{\partial c_\beta}{\partial\theta_l}\frac{\partial c_\delta}{\partial\theta_j}\right]\delta\Psi_{\alpha\beta}.
\eeq
When the error in the precision matrix arises from the estimating the sample covariance with too few mocks, the covariance of the extra parameter shift can be shown to be
\beq
    \av{\Delta\theta_i\Delta\theta_j} = \frac{(N_\mathrm{mock}-N_\mathrm{SV})(N_\mathrm{SV}-N_\mathrm{param})}{(N_\mathrm{mock}-N_\mathrm{SV}-1)(N_\mathrm{mock}-N_\mathrm{SV}-4)}\Phi_{ij}
\eeq
\citep{2013PhRvD..88f3537D}, where $N_\mathrm{param}$ is the length of the parameter vector.\footnote{Note that this includes the Hartlap factor required to invert the noisy covariance without bias.} This just inflates the usual parameter estimate by a constant factor, which, for $N_\mathrm{mock}\gg N_\mathrm{SV}$ is well approximated by $(N_\mathrm{SV}-N_\mathrm{param})/N_\mathrm{mock}$.

\section{Analytic Marginalization of the Power Spectrum Likelihood}\label{appen: an-marg}
For efficient sampling of high-dimensional parameter spaces, it is useful to marginalize over nuisance parameters analytically, as in Refs.\,\citep{2002MNRAS.335.1193B,2010MNRAS.408..865T}. Whilst one may perform this approximately for any parameters, those which enter the likelihood quadratically can be marginalized exactly, given some choice of prior, \resub{assuming that the likelihood is Gaussian}. Here, we consider the log-likelihood of Eq.\,\ref{eq: chi2-data} for parameters $\vec\theta = \{\vec\psi,\vec\eta\}$, where nuisance paramaeters $\vec\eta$ enter the model $c(\vec\theta)$ linearly. As shown in Ref.\,\citep{2010MNRAS.408..865T}, marginalizing over $\vec\eta$, assuming Gaussian priors, we obtain
\beq
     \hat{\chi}^2_{D,M}(\vec\psi) = \sum_{\alpha\beta} \left(\hat{c}_\alpha-c_\alpha(\vec\psi)\right)\mathcal{C}^{-1}_{D,M,\alpha\beta}(\vec\psi)\left(\hat{c}_\beta-c_\beta(\vec\psi)\right) + \log\left|\mathcal{C}_{D,M}(\vec\psi)\right|
\eeq
where the model is evaluated at the mean values of $\vec\eta$. Now the covariance matrix inherits dependence on cosmology via
\beq\label{eq: marg-cov}
    \mathcal{C}_{D,M,\alpha\beta}(\vec\psi) = \mathcal{C}_{D,\alpha\beta} + \sum_{ij}\frac{\partial c_\alpha(\vec\psi)}{\partial\eta_i}\frac{\partial c_\beta(\vec\psi)}{\partial\eta_j}\mathsf{P}_{ij},
\eeq
where $\mathsf{P}_{ij}$ is the prior parameter covariance (usually diagonal) and $i,j$ run over the elements of $\vec\eta$.

In our context, the counterterms $c_{s,0}$, $c_{s,2}$, $b_4$ and $P_\mathrm{shot}$ enter the power spectrum model linearly (and thus also the window convolved $P(k)$ and rotated $c(\vec\theta)$ coefficients). Analytic marginalization is possible via the methods above, using the broad Gaussian parameter priors of Ref.\,\citep{2020JCAP...05..042I}. Whilst the likelihood becomes more complex in this case (and requires additional window function derivatives for the parameter derivatives), it is still fast to compute and the parameter space can be reduced from $10$ to $6$ elements.

In some scenarios, one has a parameter, say $\zeta$, that is additionally constrained to be positive (for example the shot-noise, if an estimate has not already been subtracted). Analytic marginalization can still be performed in this case, though the expression for the log-likelihood is more complex;
\beq
     \left.\hat{\chi}^2_{D,M}(\vec\psi)\right|_{\zeta>0} &=& \sum_{\alpha\beta} \left(\hat{c}_\alpha-c_\alpha(\vec\psi)\right)\widetilde{\mathcal{C}}_{D,M,\alpha\beta}(\vec\psi)\left(\hat{c}_\beta-c_\beta(\vec\psi)\right) + \log\left|\widetilde{\mathcal{C}}_{D,M}(\vec\psi)\right|\\\nonumber
     &&\,-2\log\left[1+\mathrm{erf}\left(\sqrt{\frac{J_1(\psi)+\sigma^{-2}_{\zeta}}{2}}\right)\left(\frac{J_2(\psi)}{J_1(\psi)+\sigma_{\zeta}^{-2}}+\overline{\zeta}\right)\right]
\eeq
where $\widetilde{C}_{D,M}$ is as Eq.\,\ref{eq: marg-cov} including $\zeta$ in the nuisance parameters $\vec\eta$ and we define
\beq
    J_1(\vec\psi) &=& \sum_{\alpha\beta}\frac{\partial c_\alpha(\vec\psi)}{\partial \zeta}\frac{\partial c_\beta(\vec\psi)}{\partial \zeta}\mathcal{C}^{-1}_{D,M,\alpha\beta}\\\nonumber
    J_2(\vec\psi) &=& \sum_{\alpha\beta}\frac{\partial c_\alpha(\vec\psi)}{\partial \zeta}\mathcal{C}^{-1}_{D,M,\alpha\beta}\left(\hat{c}_\beta - c_\beta(\vec\psi)\right)
\eeq
where $\overline{\zeta}$ and $\sigma_{\zeta}$ define the Gaussian prior on the positive parameter and $\mathcal{C}_{D,M}$ does \textit{not} include $\zeta$. 

\bibliographystyle{JHEP}
\bibliography{adslib,otherlib}

\end{document}